%% file: main.tex
\documentclass[acmsmall,screen]{acmart}
\AtBeginDocument{%
  }

\setcopyright{acmcopyright}
\copyrightyear{2025}
\acmYear{2025}
\acmDOI{XXXXXXX.XXXXXXX}
\usepackage{xcolor}
\input{common_setting}

\begin{document}
\title{\input{components/title}}
\input{components/authors}

\input{components/abstract}

\input{sections/sec0_ccs}
\maketitle
\input{sections/sec1_introduction}
\input{sections/sec2_preliminaries}
\input{sections/sec3_datasetConstruction}
\input{sections/sec4_empiricalStudy}
\input{sections/sec5_threatsValidity}
\input{sections/sec6_discussion}
\input{sections/sec7_relatedwork}
\input{sections/sec8_conclusion}

\input{sections/sec9_dataCodeAvailability}
\input{sections/sec10_acknowledgments}
\bibliographystyle{ACM-Reference-Format}
\bibliography{sections/sec0_ref}
\end{document}

%% file: common_setting.tex
\usepackage{ifthen}
\newboolean{showcomments}
\setboolean{showcomments}{true} 

\ifthenelse{\boolean{showcomments}}{
	\newcommand{\nbc}[3]{
		{\colorbox{#3}{\bfseries\sffamily\scriptsize\textcolor{white}{#1}}}
		{\textcolor{#3}{\sf\small$\langle$\textit{#2}$\rangle$}}}
	
}{
	\newcommand{\nbc}[3]{}

}

\newcommand{\InlineCode}[1]{``\textsf{#1}''}

\usepackage{xspace}
\newcommand{\Category}[1]{``\emph{#1}''}
\newcommand{\Quantum}{\Category{quantum}\xspace}
\newcommand{\NotQuantum}{\Category{non-quantum}\xspace}
\newcommand{\QuantumMath}{\Category{mathematics-for-quantum}\xspace}
\newcommand{\QuantumAlgorithm}{\Category{quantum-algorithm}\xspace}
\newcommand{\QuantumCircuit}{\Category{quantum-circuit-model}\xspace}
\newcommand{\QuantumHardware}{\Category{quantum-hardware}\xspace}
\newcommand{\QuantumSoftware}{\Category{quantum-software-stack}\xspace}
\newcommand{\QuantumOthers}{\Category{quantum-others}\xspace}
\newcommand{\DevWhat}{\Category{what}\xspace}
\newcommand{\DevWhy}{\Category{why}\xspace}
\newcommand{\DevDone}{\Category{how-it-is-done}\xspace}
\newcommand{\DevUse}{\Category{how-to-use}\xspace}
\newcommand{\DevProperty}{\Category{property}\xspace}
\newcommand{\DevOthers}{\Category{developer-others}\xspace}

\newcommand{\summary}[1]{\textbf{\textit{#1}}.}

\usepackage{ifthen}
\usepackage{hyperref}
\usepackage{ifthen}
\usepackage{hyperref}
\newcommand{\RQDistributionContent}[1]{%
    \ifnum#1=0
        Structure-based perspective%
    \else\ifnum#1=1 
        What is the distribution of comments with respect to different code entities and comment forms?%
    \else\ifnum#1=2 
        How much effort have developers made to write comments for different code entities?%
    \else\ifnum#1=3 
        What is the distribution of quantum-specific comments with respect to different code entities and comment forms?%
    \fi\fi\fi\fi
}
\newcommand{\RQDistributionIndex}[1]{RQ\the\numexpr 0 + #1 \relax}
\newcommand{\RQDistributionSect}[1]{%
  \texorpdfstring{%
    \RQDistributionIndex{#1}: \RQDistributionContent{#1}%
  }{%
    \RQDistributionIndex{#1}: \RQDistributionContent{#1}%
  }%
}
\newcommand{\RQDeveloperContent}[1]{%
    \ifnum#1=0
        Developer-intent perspective%
    \else\ifnum#1=1 
        How do developers' intentions expressed in code comments differ between QSE and CSE\mbox{}?%
    \else\ifnum#1=2 
        What are developers' intentions in code comments specific to QSE\mbox{}?%
    \fi\fi\fi%
}
\newcommand{\RQDeveloperIndex}[1]{RQ\the\numexpr 3 + #1 \relax}
\newcommand{\RQDeveloperSect}[1]{%
  \texorpdfstring{%
    \RQDeveloperIndex{#1}: \RQDeveloperContent{#1}%
  }{%
    \RQDeveloperIndex{#1}: \RQDeveloperContent{#1}%
  }%
}
\newcommand{\RQQuantumContent}[1]{%
    \ifnum#1=0
        Quantum-specific perspective%
    \else\ifnum#1=1 
        What is the distribution of code comments across quantum-specific categories?%
    \else\ifnum#1=2 
        What is the relationship between developer-intent and quantum-specific taxonomies in code comments?%
    \else\ifnum#1=3 
        What semantic differences among quantum-specific categories are revealed through code comments?%
    \fi\fi\fi\fi%
}
\newcommand{\RQQuantumIndex}[1]{RQ\the\numexpr 5 + #1 \relax}
\newcommand{\RQQuantumSect}[1]{%
  \texorpdfstring{%
    \RQQuantumIndex{#1}: \RQQuantumContent{#1}%
  }{%
    \RQQuantumIndex{#1}: \RQQuantumContent{#1}%
  }%
}

\usepackage{xcolor}
\usepackage[most]{tcolorbox}

\newcommand{\AnswertoRQ}[2]{
  \begin{tcolorbox}[
    colback=black!1!white,
    colframe=black!40!white,
    left=0.6mm, 
    right=0.6mm, 
    top=0.6mm, 
    bottom=0.6mm, 
    boxsep=0.8mm, 
    arc=2mm,
  ]
  \textbf{Findings of #1:} #2
  \end{tcolorbox}
}

\usepackage[normalem]{ulem}

\usepackage{graphicx}
\usepackage{subcaption} 
\usepackage{enumitem}  

\usepackage{array}
\usepackage{setspace}
\usepackage{ragged2e} 
\newcolumntype{P}[1]{>{\justifying\arraybackslash\setstretch{0.5}\parindent=0pt}p{#1}}

\usepackage{braket}

\usepackage[utf8]{inputenc}
\usepackage{geometry}

\usepackage{listings}
\usepackage{xcolor}
\usepackage{adjustbox}
\usepackage{float} 
\usepackage{mathrsfs}

\lstdefinestyle{mypython}{
    language=Python,
    basicstyle=\fontfamily{cmss}\fontsize{7pt}{7pt}\selectfont,
    columns=fixed,
    keywordstyle=\color{blue}\bfseries,
    commentstyle=\color{green!50!black}\itshape,
    stringstyle=\color{red},
    breaklines=true,
    breakatwhitespace=true,
    numbers=left,
    numberstyle=\tiny,
    numbersep=5pt,
    captionpos=b 
}

\usepackage{pifont}
\newcommand{\y}{\ding{51}}
\newcommand{\n}{\ding{55}}

%% file: components/title.tex
Code Comments for Quantum Software Development Kits: An Empirical Study on Qiskit

%% file: components/authors.tex
\author{Zenghui Zhou}
\authornote{The two authors contributed equally to this work.}
\orcid{0000-0002-1824-6979}
\email{zhouzenghui@buaa.edu.cn}
\affiliation{%
    \institution{Beihang University}
    \city{Beijing}
    \country{China}
}

\author{Yuechen Li}
\authornotemark[1]
\orcid{0009-0006-5109-3288}
\email{liyuechen@buaa.edu.cn}
\affiliation{%
    \institution{Beihang University}
    \city{Beijing}
    \country{China}
}

\author{Yi Cai}
\orcid{0009-0000-2190-0030}
\email{caiy@buaa.edu.cn}
\affiliation{%
    \institution{Beihang University}
    \city{Beijing}
    \country{China}
}

\author{Jinlong Wen}
\orcid{0009-0009-2027-183X}
\email{wjloong@buaa.edu.cn}
\affiliation{%
    \institution{Beihang University}
    \city{Beijing}
    \country{China}
}

\author{Xiaohan Yu}
\orcid{0009-0003-4595-7359}
\email{xxxiyxh@buaa.edu.cn}
\affiliation{%
    \institution{Beihang University}
    \city{Beijing}
    \country{China}
}

\author{Zheng Zheng}
\authornote{Corresponding author.}
\orcid{0000-0001-7922-9067}
\email{zhengz@buaa.edu.cn}
\affiliation{%
    \institution{Beihang University}
    \city{Beijing}
    \country{China}
}

\author{Beibei Yin}
\orcid{0000-0002-8938-1203}
\email{yinbeibei@buaa.edu.cn}
\affiliation{%
    \institution{Beihang University}
    \city{Beijing}
    \country{China}
}

%% file: components/abstract.tex
\begin{abstract}
Quantum computing is gaining attention from academia and industry. With the quantum Software Development Kits (SDKs), programmers can develop quantum software to explore the power of quantum computing. However, programmers may face challenges in understanding quantum software due to the non-intuitive quantum mechanics. To facilitate software development and maintenance, code comments offered in quantum SDKs serve as a natural language explanation of program functionalities and logical flows. Despite their importance, scarce research systematically reports their value and provides constructive guidelines for programmers. To address this gap, our paper focuses on Qiskit, one of the most popular quantum SDKs, and presents CC4Q, the first dataset of code comments for quantum computing. CC4Q incorporates 9677 code comment pairs and 21970 sentence-level code comment units, the latter of which involve heavy human annotation. Regarding the annotation, we validate the applicability of the developer-intent taxonomy used in classical programs, and also propose a new taxonomy considering quantum-specific knowledge. We conduct an empirical study comprehensively interpreting code comments from three perspectives: comment structure and coverage, developers' intentions, and associated quantum topics. Our findings uncover key differences in code comments between classical and quantum software, and also outline quantum-specific knowledge relevant to quantum software development.
\end{abstract}

%% file: sections/sec0_ccs.tex
\ccsdesc[300]{Software and its engineering~Software development techniques}
\ccsdesc[300]{Computer systems organization~Quantum computing}

%% file: components/keywords.tex
quantum software engineering, code comments, software development kits, empirical study

%% file: sections/sec1_introduction.tex
\section{Introduction}

Quantum computing is a promising technique to solve complex computational problems. Due to the unique characteristics of quantum mechanics, such as superposition, entanglement, and quantum tunneling, quantum computers can process multiple inputs of a function simultaneously. Compared to classical computing using bits and logical gates, quantum computing employs qubits (a.k.a. quantum bits) as the basic computational unit, and utilizes quantum gates for operations in gate-based systems like trapped ion and superconducting quantum computers. Nowadays, many quantum \textbf{Software Development Kits (SDKs)}, such as IBM Qiskit~\cite{aleksandrowicz2019qiskit}, Google Cirq~\cite{google2018cirq}, and Microsoft Quantum Development Kit~\cite{svore2018q}, are available to the public, enabling developers to start quantum programming and have access to real quantum hardware or its classical simulation. As one of the most widely used quantum SDKs revealed in recent studies~\cite{zhang2025quantum, IshimotoNakamura2024}, Qiskit is built on the host language Python and has made significant contributions to the quantum computing ecosystem in terms of development and accessibility. Since human intuition aligns better with the classical world than quantum mechanics~\cite{piattini2021toward, ying2012floyd}, developers are prone to making mistakes when writing quantum programs, thereby potentially threatening the quality of quantum programs.

From the viewpoint of software development and maintenance, a feasible solution is to provide readable and valid documentation for developers to comprehend the functions and logical flows of the programs. In this way, program bugs stemming from the developers' limited expertise in quantum computing can be somewhat mitigated. In software engineering, code comments, as one form of code documentation, are presented in natural language and designed to describe both the functionality and rationale of the code. Regarding quantum programs, the included code comments can contain rich and readable texts for developers, capable of explaining unfamiliar terms specific to quantum physics and quantum computing. Take an instance of a comment in Qiskit, ~\InlineCode{Perform a controlled-X gate on qubit 1, controlled by qubit 0}, which corresponds to the code snippet~\InlineCode{qc.cx(0, 1)}. It appears tough for developers to gain more details merely from such a short code snippet. The code comment clearly depicts a principle that a qubit controls the operation on the other qubit in this specific gate. With this code comment, developers can learn a vital property of controlled operations in quantum computing and distinguish the two different input arguments when programming.

\textbf{Quantum Software Engineering (QSE)} is an emerging research field with a focus on the principles, methodologies, standards, and tools for quantum software development life cycle~\cite{zhao2020quantum, murillo2025quantum, pezze20252030}. 
Until now, code documentation for quantum computing has received scarce attention in QSE. 
%
Only two publications~\cite{d2024exploring, YuLi2025} investigated general code explanation for quantum computing, where these works leverage large language models to generate code summarizations and code comments for quantum programs, respectively. Due to the absence of a public benchmark for code comments in quantum SDKs, the evaluation of generated code comments in the two works requires human involvement, thereby incurring extra cost and subjectivity. Furthermore, in the view of software development, many developers only pay attention to the code and ignore the comments and documentation~\cite{yang2019survey}. If this practice is also applied in the development of quantum software, its readability and maintainability could be especially undermined. Indeed, developers can be trained to understand and write code comments by instances from high-quality and well-documented quantum software repositories. Nevertheless, code comments in such repositories (e.g., those of Qiskit) have not been reported sufficiently, resulting in a lack of guidelines for developers.
 
To help software developers understand and engage with quantum software better, our paper presents CC4Q, the first dataset of code comments for quantum SDKs. CC4Q includes exemplary comment instances from Qiskit, accompanied by detailed annotated data for our empirical analysis. In detail, CC4Q collects 9,677 \textbf{Code Comment Pairs (CCPs)} originally from a core component library of Qiskit, and incorporates 21,970 \textbf{Sentence-level Code Comment Units (SCCUs)} to facilitate textual comprehension and data analysis, where the SCCUs are derived by reasonably segmenting CCPs. Since not all the code comments in segmented SCCUs are associated with quantum physics or quantum computing, each SCCU is first examined for its relevance to quantum topics. In detail, an SCCU is marked as \Quantum iff it is significantly quantum-specific; otherwise, it turns out to be \NotQuantum. In assessing whether techniques of code comments in \textbf{Classical Software Engineering (CSE)} can be adapted to QSE, we validate the applicability of the developer-intent taxonomy~\cite{chen2021my} proposed for classical Java programs to Python programs for quantum computing. We then label each SCCU based on the developer-intent categories (e.g., \DevWhat, \DevWhy). To further assist developers in learning the software landscape of quantum computing, our paper presents a new taxonomy (named quantum-specific taxonomy) as a fine-grained classification of \Quantum SCCUs according to the described quantum topics. Categories (e.g., \QuantumMath, \QuantumAlgorithm) in this taxonomy are determined by quantum-specific knowledge conveyed in code comments. Generally, we manually annotate SCCUs for nearly one month, while also exploring deep learning models to infer the remaining SCCUs based on the manually labeled data. Each SCCU in CC4Q encompasses the three types of categories mentioned above.


After constructing CC4Q, we conduct an empirical study including eight research questions to interpret code comments from three distinct perspectives, i.e., (1) structure-based perspective: investigating the distribution of comments across code entities (i.e., function, class, and module) and forms (i.e., docstring, block comment, and inline comment) to pinpoint where developers concentrate their documentation efforts and which forms best convey quantum complexity; (2) developer-intent perspective: analyzing underlying developers' intention for quantum software, and the distinction between developing quantum software and classical software in this aspect; and (3) quantum-specific perspective: focusing on domain topics to help more developers understand the quantum-specific content and knowledge.
Overall, the results provide a comprehensive overview of official code comments in Qiskit, including their forms, associated code entities, and the quantum-specific knowledge they contain. Based on these findings, we offer practical guidelines for developers to create higher-quality comments for quantum programs.

In general, we summarize three main contributions of our paper: 

\begin{itemize} 
    \item We present the first comprehensive dataset of code comments for QSE and the artifact is publicly available at~\cite{cc4q2025artifact}.
    \item We propose a novel taxonomy tailored for code comments specific to quantum physics and quantum computing.
    \item We conduct a comprehensive empirical study based on the dataset and provide insights for code comments when developing programs with quantum SDKs.
\end{itemize}

The rest of our paper is organized as follows. Section~\ref{sec:preliminaries} provides preliminary knowledge involved in our paper. Section~\ref{sec:method} elaborates on the method of constructing CC4Q. Empirical results and analysis are demonstrated in Section~\ref{sec:empirical}, followed by threats to validity in~\ref{sec:threats}. In Section~\ref{sec:discussion}, we discuss our findings and propose reasonable guidelines for developers. Section~\ref{sec:related} compares our work with related work. Finally, Section~\ref{sec:conclusion} concludes our paper and outlines future directions. 

%% file: sections/sec2_preliminaries.tex
\section{Preliminaries}
\label{sec:preliminaries}

\subsection{Quantum Information and Quantum Computing}
In quantum computing, a qubit is the basic computational unit that can represent the superposition of two basis states. For instance, a single-qubit state can be written in the form of a linear combination, i.e., $\ket{\psi}=\alpha\ket{0}+\beta\ket{1}$, where the amplitudes $\alpha$ and $\beta$ are complex numbers. Two or more qubits can form an entangled state in which no component can be described independently of the others. For gate-based quantum systems, quantum gates govern the evolution of qubits, enabling them to be processed in a way analogous to logic operations in classical computing. There are many basic gates provided in quantum SDKs (e.g., Pauli-X gate, Hadamard gate, and controlled-NOT gate). Unlike classical logic gates, quantum gates are mathematically unitary and reversible, in accordance with the postulates of quantum mechanics~\cite{nielsen2010quantum}. Quantum measurement is the only means of extracting computational results from qubits, and the outcomes are inherently probabilistic, like the probability of obtaining a $0$ from $\ket{\psi}$ as $|\alpha|^2$.

Regarding the design of quantum programs, quantum circuits serve as a model to visualize the procedure of a quantum algorithm. A complete quantum algorithm consists of encoding of data, a sequence of quantum gates applied to the input qubits, and measurement of at least one qubit at the end~\cite{jayakumar2022quantum}. The quantum software stack offers a multi-layer software architecture that bridges the gap between abstract design and physical quantum devices, where the user interface, programming language, compilers, backends, etc., exist at different layers in the software stack~\cite{serrano2022quantum, wang2021qdiff}. At the top, highly abstracted \textbf{Application Programming Interfaces (APIs)} offered by quantum SDKs allow developers to construct, simulate, and optimize quantum circuits. Note that current quantum computers are in the \textbf{Noisy Intermediate-Scale Quantum (NISQ)} era, where real quantum hardware can only equip middle-scale qubits and involve noise caused by factors like decoherence and crosstalk. Thus, quantum software engineers should perform the circuit mapping and transpilation through SDK-provided APIs to match the target real quantum hardware.

\subsection{Comment Forms for Python Code}
A code comment is the text embedded in source code to make the code easier for a programmer to understand. Nowadays, the practice of writing code comments has become standardized, particularly with regard to their presented forms. Taking an instance of Python code, PEP 8-style guide recommends that comments are complete sentences and presented in English~\cite{van2001pep}. Furthermore, there are three regular forms of comments for Python code, i.e., inline comments, block comments, and documentation strings (a.k.a., docstrings). We show their examples in Listings~\ref{lst:inline}-\ref{lst:docstring}, where the code snippets are picked from our dataset CC4Q. An inline comment should be separated by at least two spaces from the statement and start with \InlineCode{\#} followed by a single space. This type of comment is located on the same line as the code statement to be explained. In comparison, a block comment generally applies to multiple lines of code, and each line of a block comment looks similar to a single inline comment. Docstrings are usually written for all public modules, functions, classes, and methods. Unlike inline and block comments, docstrings are preserved in the abstract syntax tree and retained throughout the runtime of the program. According to the docstring conventions~\cite{goodger2001pep}, docstrings can be presented in single or multiple lines, and both are suggested to start and end with \InlineCode{"""}.

\begin{table}[!t]
    \centering
    \input{tables/code_form_examples}
\end{table}

 

%% file: tables/code_form_examples.tex
\begin{center}
\begin{adjustbox}{max width=\textwidth}
\begin{tabular}[!t]{c}
\begin{minipage}[!t]{0.9\textwidth}
\begin{lstlisting}[style=mypython, caption={The inline comment corresponding to the statement \InlineCode{self.leader = \{\}}}, label={lst:inline}]
def __init__(self):        
    self.leader = {}  # qubit's group leader
    self.group = {}
\end{lstlisting}
\end{minipage}
 \\
\begin{minipage}[!t]{0.9\textwidth}
\lstset{style=mypython}
\begin{lstlisting}[style=mypython, caption={The block comment describing the function of the \InlineCode{for}-loop structure}, label={lst:block}]
def collapse_to_operation(self, blocks, collapse_fn):
    ...
    for block in blocks:
        # Find the sets of qubits/clbits used in this block (which might be much smaller
        # than the set of all qubits/clbits).
        cur_qubits = set()
        cur_clbits = set()
        cur_clregs = set()
        ...
\end{lstlisting}
\end{minipage}
 
\\
\begin{minipage}[!t]{0.9\textwidth}
 
\begin{lstlisting}[style=mypython, caption={The docstring explaining the function \InlineCode{reverse\_qargs}}, showspaces=false, showstringspaces=false, label={lst:docstring}]
def reverse_qargs(self) -> Statevector:
    r"""Return a Statevector with reversed subsystem ordering.

    For a tensor product state this is equivalent to reversing the order
    of tensor product subsystems. For a statevector
    :math:`|\psi \rangle = |\psi_{n-1} \rangle \otimes ... \otimes |\psi_0 \rangle`
    the returned statevector will be
    :math:`|\psi_{0} \rangle \otimes ... \otimes |\psi_{n-1} \rangle`.

    Returns:
        Statevector: the Statevector with reversed subsystem order.
    """
    ret = _copy.copy(self)
    ...
\end{lstlisting}
\end{minipage}
\end{tabular}
\end{adjustbox}
\end{center}

%% file: sections/sec3_datasetConstruction.tex
\section{Method for Dataset Construction}
\label{sec:method}
In this section, we introduce the method to construct our dataset CC4Q, along with the involved data profile.

\subsection{Procedure of Data Collection and Pre-processing}
To construct our dataset, we select a core component library of Qiskit\footnote{\url{https://github.com/Qiskit/qiskit/tree/main}} as the source of code comments. Our consideration is motivated by the widespread adoption of this open-source quantum SDK in QSE, particularly for areas like quantum software testing~\cite{leite2025testing} and quantum optimization for software engineering activities~\cite{zhang2025quantum}. Besides, the object library is designed for general-purpose quantum experience, making it accessible to a broad spectrum of practitioners across diverse domains. 


We use LibCST\footnote{\url{https://github.com/Instagram/LibCST}} (Library for Concrete Syntax Trees) to parse the source files, which allows us to extract CCPs effectively by preserving comments and formatting information. For each CCP, we first record its raw data, i.e., the code comment and its associated code snippet. Next, we automatically generate several labels showing important properties of a CCP, including its comment form and the enclosing code entity. A detailed description of all recorded terms within a CCP will be provided in Section~\ref{sec:CCP-structure}.

Following Zhai et al.~\cite{zhai2020cpc}, we segment each CCP into SCCUs as well. This choice is motivated by the fact that multi-line comments often contain heterogeneous information, such as describing the functionality of a function and detailing its implementation. Hence, treating sentences as the basic unit for analysis allows us to capture the diverse patterns within a long comment in quantum SDKs, and enables the finer-grained comparison across comment forms and code entities as well. Each SCCU inherits the metadata of its original CCP, ensuring that traceability to the source code is preserved. Technically, the segmentation in our paper uses Docutils\footnote{\url{https://sourceforge.net/projects/docutils}} and spaCy\footnote{\url{https://github.com/explosion/spaCy}}, where Docutils helps parse reStructuredText formatting commonly used in Python docstrings, and spaCy provides reliable sentence segmentation.
In total, our dataset CC4Q incorporates 9,677 CCPs along with 21,970 segmented SCCUs from these CCPs.

\subsection{Overview of Annotation Tasks and Multi-class Taxonomies}\label{sec:annotation-tasks}

This subsection introduces three annotation tasks that support our labeling and empirical analysis. 
They include a binary classification task for identifying quantum-specific comments (\textbf{TASK1}), a multi-classification task for recognizing developers' intentions with existing categories (\textbf{TASK2}), and a multi-classification task for extracting involved quantum topics with categories proposed in our paper (\textbf{TASK3}).


\begin{table}[!t]
    \small
    \centering
    \caption{Descriptions and examples for the developer-intent taxonomy}
    \label{tab:developer-taxonomy}
    \resizebox{.98\textwidth}{!}{
        \input{tables/developer-intent}
    }
\end{table}

\subsubsection{\textbf{TASK1}: Identification of Quantum-specific Comments} 
\label{sec:boolean}
This task is motivated by an initial inquiry into the proportion of code comments in quantum SDKs that describe terms specific to quantum. In detail, we define that a comment is marked as \Quantum\ only if it is \uline{significantly} \uline{quantum}-\uline{specific}; otherwise, it is not \Quantum (marked as \NotQuantum in our paper). That is to say, \textbf{TASK1} merely aims to yield the Boolean labels. Herein, the term \uline{quantum} is specified to principles, concepts, techniques, or phenomena in quantum physics or quantum computing. The word \uline{specific} is highlighted to exclude notions shared with classical computing. For example, a comment that only describes the role of an unspecified register is not \Quantum, since registers are ubiquitous in classical digital circuits as well. At last, the adverb \uline{significantly} points out direct and explicit relevance rather than indirect or ambiguous mentions.

\subsubsection{\textbf{TASK2}: Classification for Developer-intent Taxonomy}
\label{sec:developer-intent}
This taxonomy classifies each SCCU by various intentions of the developers to examine the adaptability of the taxonomy used in a prior study within CSE~\cite{zhai2020cpc}. We fully reuse five categories shared in that work, where Table~\ref {tab:developer-taxonomy} presents their definitions and comment examples in Qiskit. These candidates span functional summaries (\DevWhat), rationales (\DevWhy), implementation details (\DevDone), usage guidelines (\DevUse), and behavioral constraints (\DevProperty). During our pilot coding (as further introduced in Section~\ref{sec:human}), we observed that a portion of the comments could not be consistently assigned to any of the existing five categories. For instance, academic references are clearly presented in Qiskit and have educational or traceable purposes. To this end, we additionally consider a category, i.e., \DevOthers, which contains these cases hardly attributed to any of the above five categories to ensure the intuitive completeness of this developer-intent taxonomy. We remark that our defined \DevOthers is different from the category \Category{others} in two existing papers~\cite{mu2023developer, chen2021my}, the latter category that refers to unspecified or ambiguous comments.




\subsubsection{\textbf{TASK3}: Classification for Quantum-specific Taxonomy}
\begin{table*}[!t]
    \small
    \centering
    \caption{Descriptions and examples for the quantum-specific taxonomy}
    \label{tab:quantum-taxonomy}
    \resizebox{.98\textwidth}{!}{
        \input{tables/quantum-specific}
    }
\end{table*}

This task is proposed to capture domain-specific content in the code comments provided by quantum SDKs. By providing an overview of quantum-specific knowledge, this is promising to help software engineers engaged in quantum software development to deepen their understanding of quantum physics and quantum computing. For this task, we concentrate on SCCUs categorized as \Quantum, and propose the quantum-specific taxonomy with six categories tailored to quantum SDKs upon the coding procedure. Table~\ref{tab:quantum-taxonomy} summarizes their formal definitions and representative examples. The first five categories correspond to different topics well defined, while \QuantumOthers accommodates \Quantum comments vague or ambiguous to be regarded within other quantum topics. Among these categories, we emphasize that \QuantumAlgorithm captures high-level and systematic descriptions of specific quantum algorithms, compared to \QuantumCircuit more oriented toward the details of ordinary quantum circuits. Unlike conventional perceptions of software and code, \QuantumSoftware underscores a multi-layer software architecture, such as the assembly-based programming language OpenQASM at the layer of quantum application~\cite{serrano2022quantum}. Moreover, \QuantumHardware is vital in QSE, an interdisciplinary domain that naturally accounts for constraints and attributes of physical quantum computers.


\subsection{Details of Implementing Comment Annotation}
\label{ec:annotation-procedure}

\begin{table}[!t]
    \small
    \centering
    \caption{Summary of comment annotation details for the three tasks.}
    \label{tab:annotation}
    \resizebox{.98\textwidth}{!}{
        \input{tables/performance_of_3_models}

    }
    {\justify
        The terms ``Keywords?'' and ``Coding?'' respectively indicate whether keyword matching and open coding are involved. In ``\# of SCCUs'', we mark the number of SCCUs actually labeled by annotators with a parenthesis.
    \par}
\end{table}

We employ a human-in-the-loop pipeline to generate labels for the three tasks. Three tasks are completed from \textbf{TASK1} to \textbf{TASK3} in order. For each task, we
manually annotate almost 10\% of SCCUs in CC4Q, use these annotations as ground truth, and train
deep learning classifiers to infer the labels of the remaining SCCUs. Table~\ref{tab:annotation} summarizes the details involved in implementing comment annotation.

\subsubsection{Human Annotation}
\label{sec:human}

For the phase of human annotation, three Ph.D. students and two master’s students annotate SCCUs independently. Among the five annotators, the pilot annotator is a Ph.D. student with three years of research experience in QSE. The other two Ph.D. students have more than three years of experience in software engineering and are familiar with Python. The two master’s students have been working with Qiskit for at least one year. Before implementing the comment annotation, the pilot annotator trains the other four with the preliminaries of quantum computing and delivers detailed guidelines for each task. 

Especially, we employ the scheme of keyword matching for \textbf{TASK1} to reduce the human cost and ensure the annotation quality. This is because a comment is definitely \Quantum once it includes obviously quantum-specific terms like ``quantum'' and ``superposition'', where the keyword list is summarized by annotators and shared in our artifact. Consequently, keyword matching effectively identified 333 \Quantum SCCUs free of human annotation. However, keyword matching is not applied to \textbf{TASK2} and \textbf{TASK3}, as the two tasks demand more in-depth semantics for the non-binary classifications, instead of simple keywords. Following Zhai et al.~\cite{zhai2020cpc}, an open coding procedure is led by the pilot annotator to ensure the rationality of the adopted multiple categories in \textbf{TASK2} and \textbf{TASK3} before human annotation. After finishing the annotation of \textbf{TASK1}, the pilot annotator recorded personal interpretation of each of the randomly sampled 300 \Quantum and 300 \NotQuantum SCCUs, focusing on the developers’ intentions and possible quantum-specific knowledge involved. As a result of open coding, we validated the reasonability of adapting developer-intent categories to classify code comments in quantum SDKs, and also summarized the six candidate categories in our proposed quantum-specific taxonomy.

SCCUs in each task are randomly sampled from the corresponding data pools for human annotation, during which no fewer than 2,000 SCCUs must be categorized per task. Each category of an SCCU is labeled by two annotators, and the inter-rater agreement is measured using Cohen's kappa coefficient ($\kappa$)~\cite{pontius2011death}. The $\kappa$ coefficients are listed in Table~\ref{tab:annotation}, and we can conclude that substantial agreement has been achieved for three tasks, according to the consistency level suggested in~\cite{landis1977measurement}. Even so, we resolve all the resulting disagreements through group discussion and retain a single category by consensus for each task.

\subsubsection{Model Training and Inference}
In consideration of cost-effectiveness, we train a BERT-based classifier per task and use the yielding models to infer categories for the remaining unlabeled SCCUs. We adopt an 8:1:1 split of the human-annotated data for training, validation, and testing. BERT is selected due to its strong capability in representing natural language texts and its proven effectiveness in classification tasks~\cite{devlin2019bert}. 
Table~\ref{tab:annotation} presents the performance of our trained models by Accuracy, Precision, Recall, and F1-score calculated for the test sets. All the scores are above 0.850, suggesting the acceptable quality of the automatic annotation to some extent.

\subsection{Profile of Data Formats}\label{sec:CCP-structure}

In the constructed CC4Q, each CCP is assigned a unique ID, and incorporates both textual items of a code comment and the corresponding code snippet. Apart from those, we include some basic information beneficial for management and maintenance, such as the associated SDK, the software version, the host language, the file directory, and the range of code lines. Specifically, we automatically extract the structural attributes for each CCP, including the code entity (i.e., module, class, or function) and comment form (i.e., inline, block, or docstring). The identification of comment forms obeys the standard conventions of Python comments~\cite{van2001pep}. Concerning code entities, the term function refers to both module-level functions and class-level methods, following the extraction granularity of LibCST. We do not treat statements as a separate entity type, since comments cannot always be reliably mapped to individual statements. Inline and block comments that appear inside functions are therefore associated with the enclosing function.

Each SCCU inherits the metadata of its parent CCP, while a new ID is assigned with the same prefix as the original CCP ID. The comment text of an SCCU corresponds to the segmented sentence rather than the full comment. As described in Section~\ref{sec:annotation-tasks}, the annotation result per task is included, i.e., a Boolean label for \textbf{TASK1}, one of the six developer-intent categories for \textbf{TASK2}, and one of the six quantum-specific categories for \textbf{TASK3}.

%% file: tables/developer-intent.tex
\begin{tabular}{P{0.1\textwidth}|P{0.42\textwidth}|P{0.46\textwidth}}
    \toprule[1pt]
    \multicolumn{1}{c|}{\textbf{Categories}}     & \multicolumn{1}{c|}{\textbf{Descriptions}} &\multicolumn{1}{c}{\textbf{Comment examples}} \\ 
    \midrule
    \DevWhat           
    & Provides a definition or a summary of functionality of the subject and/or its interface.  
    & \InlineCode{Checks whether op can be converted to Gate.}
    \\
    \midrule
    \DevWhy 
    & Explains the reason why the subject is provided or the design rationale of the subject.
    & \InlineCode{For legacy reasons the uuid is stored in a different format as this was done prior to QPY 11.}
    \\
    \midrule
    \DevDone     
    & Describes the implementation details of a method.
    & \InlineCode{Open controls are implemented by conjugating the control line with X gates.}
    \\
    \midrule
    \DevUse       
    & Describes the usage or the expected set-up of using a method
    & \InlineCode{This is designed to be used by both .QuantumCircuit and .DAGCircuit when managing operations that need to map classical resources from one circuit to another.}
    \\
    \midrule
    \DevProperty      
    & Asserts properties of a method, including pre-conditions or post-conditions of a method.
    & \InlineCode{This will raise a CircuitError if type safety cannot be ensured.}
    \\
    \midrule
    \DevOthers                 
    & Encompasses code comments that can not be well labeled with any of the above developer-intent categories.
    & \InlineCode{References:: Vale et. al., Circuit Decomposition of Multicontrolled Special Unitary Single-Qubit Gates, IEEE TCAD 43(3) (2024),arXiv:2302.06377}
    \\
    \bottomrule[1pt]
\end{tabular}

%% file: tables/quantum-specific.tex
\begin{tabular}{P{0.14\textwidth}|P{0.48\textwidth}|P{0.35\textwidth}}
    \toprule[1pt]
    \multicolumn{1}{c|}{\textbf{Categories}}     & \multicolumn{1}{c|}{\textbf{Descriptions}} &\multicolumn{1}{c}{\textbf{Comment examples}} \\ 
    \midrule
    \QuantumMath            
    & Includes mathematical-related concepts, expressions, or theorems usually served in quantum information theory.
    & \InlineCode{RZ(lambda)\^{}\{dagger\} = RZ(-lambda)} 
    \\
    \midrule
    \QuantumAlgorithm       
    & Discusses specific quantum algorithms, including the realization and evaluation of these quantum algorithms.
    & \InlineCode{This class uses the Shukla-Vedula algorithm [1], which only needs O(log\_2 (M)) qubits and O(log\_2 (M)) gates, to prepare the superposition.}
    \\
    \midrule
    \QuantumCircuit         
    & Zooms into design, optimization, or evaluation of quantum circuits.
    & \InlineCode{Set the global phase of the circuit.} 
    \\
    \midrule
    \QuantumHardware        
    & Mentions principles, concepts, or techniques, especially for NISQ hardware or physical implementations of models.
    & \InlineCode{construct a Layout from a bijective dictionary, mapping virtual qubits to physical qubits.}
    \\
    \midrule
    \QuantumSoftware        
    & Merely focuses on concrete principles, concepts, or techniques, particularly for the quantum software stack.
    & \InlineCode{Parse an OpenQASM 2 program from a string into a .QuantumCircuit.}
    \\
    \midrule
    \QuantumOthers                 
    & Encompasses \Quantum code comments that relate to vague and ambiguous quantum topics beyond the above five categories.
    & \InlineCode{The blocks in the entanglement layers.}
    \\
    \bottomrule[1pt]
\end{tabular}

%% file: tables/performance_of_3_models.tex
 


\begin{tabular}{l cc cc cccc}
\toprule
\textbf{Tasks}  & \multicolumn{4}{c}{\textbf{Human annotation}} & \multicolumn{4}{c}{\textbf{Model training and inference}} \\
\cmidrule(lr){2-5}\cmidrule(lr){6-9}
& \textbf{Keywords?} & \textbf{Coding?} & \textbf{\# of SCCUs} & \textbf{Cohen’s $\kappa$} & \textbf{Accuracy} & \textbf{Precision} & \textbf{Recall} & \textbf{F1-score} \\
\midrule
\textbf{TASK1} & \y & \n & 2{,}196 (1,863) & 0.739 & 0.941 & 0.956 & 0.890 & 0.922 \\
\textbf{TASK2} & \n & \y & 4{,}000 (4,000) & 0.771 & 0.864 & 0.863 & 0.865 & 0.862 \\
\textbf{TASK3} & \n & \y & 2{,}000 (2,000) & 0.808 & 0.875 & 0.875 & 0.878 & 0.876 \\
\bottomrule
\end{tabular}

%% file: sections/sec4_empiricalStudy.tex
\section{Empirical Study}
\label{sec:empirical}

\input{sections/empirical/questions}

\input{sections/empirical/distribution-based}

\input{sections/empirical/developer-intent}

\input{sections/empirical/quantum-specific}

%% file: sections/empirical/questions.tex
\subsection{Research Questions}

In our paper, we interpret code comments in Qiskit from structure-based, developer-intent, and quantum-specific perspectives. The corresponding eight \textbf{Research Questions (RQs)} are listed as follows,
\begin{itemize}[leftmargin=*]
    \item \textbf{\RQDistributionContent{0}:}
    \begin{enumerate}[leftmargin=2.5em]
        \renewcommand{\labelenumi}{\textit{\textbf{RQ\arabic{enumi}}}}
        \item \textit{\RQDistributionContent{1}}
        \item \textit{\RQDistributionContent{2}}
        \item \textit{\RQDistributionContent{3}}
    \end{enumerate}
    \item \textbf{\RQDeveloperContent{0}:}
    \begin{enumerate}[leftmargin=2.5em]
        \setcounter{enumi}{3}
        \renewcommand{\labelenumi}{\textit{\textbf{RQ\arabic{enumi}}}}
        \item \textit{\RQDeveloperContent{1}}
        \item \textit{\RQDeveloperContent{2}}
    \end{enumerate}
    \item \textbf{\textbf{\RQQuantumContent{0}:}}     
    \begin{enumerate}[leftmargin=2.5em]
        \setcounter{enumi}{5}
        \renewcommand{\labelenumi}{\textit{\textbf{RQ\arabic{enumi}}}}
        \item \textit{\RQQuantumContent{1}}    
        \item \textit{\RQQuantumContent{2}}   
        \item \textit{\RQQuantumContent{3}}   
    \end{enumerate}
\end{itemize}

The preliminary analysis from the first perspective (RQ1–RQ3) focuses on the structure and coverage of code comments, without delving deeply into semantic details. This initial perspective is motivated by the limitation of the existing work~\cite{YuLi2025} on code comments for QSE, which focused on semantics but lacked comment structures. In comparison, further analysis from the developer-intent (RQ4-RQ5) and quantum-specific (RQ6-RQ8) perspectives depends on fine-grained semantic information corresponding to our labeled two types of categories in \textbf{TASK2} and \textbf{TASK3}. The developer-intent perspective provides a fair and justified basis for comparing the contextual information of code comments for QSE with that for CSE. Specifically, by applying and extending intent taxonomies used in CSE, this perspective enables direct comparison between QSE and CSE. It also identifies the pedagogical and maintenance needs from the developer aspect, such as potentially higher demand for \DevUse comments to explain complex APIs hard to understand, as well as non-intuitive principles of quantum physics and quantum computing.  
Focusing on comments classified as \Quantum, the last perspective delves into domain content to characterize quantum-specific knowledge relevant to programming with quantum SDKs.
This is crucial for quantifying the prevalence of specific quantum knowledge (e.g., mathematics, hardware concepts). 
The insights derived from this final perspective help practitioners obtain a more comprehensive view during quantum software development.

Regarding the data support for these RQs, \RQDistributionIndex{1} and \RQDistributionIndex{2} are answered based on the collected CCPs.  Compared with SCCUs produced through extra segmentation, CCPs not only preserve the complete original comments but also provide an objective reflection of their structural distributions.
In contrast to the above two RQs involved no semantics, addressing the other RQs relies on the segmented SCCUs, owing to their better readability for sentence-level semantic analysis.  

%% file: sections/empirical/distribution-based.tex
\subsection{Empirical Results and Analysis for the Structure-based Perspective}

Analyzing the structural properties of comments is essential for uncovering their organization in quantum SDKs. To address the proposed three RQs from the structure-based perspective, we consider code entities and comment forms as two key elements reflecting the overall comment structure. Initially, RQ1 studies the frequencies of comments across different code entities and comment forms, providing an overview of documentation practices. RQ2 further investigates the comment lines to assess the proportions of developers' effort dedicated to writing comments for different entities within code snippets. RQ3 presents a particular view of comments relevant to quantum topics and also reports their frequencies in terms of different code entities and comment forms. This RQ only explores the Boolean classification mentioned in Section~\ref{sec:boolean} and serves as the cornerstone for the subsequent interpretation of comments through in-depth semantic analysis.



  


\subsubsection{\RQDistributionSect{1}}

\begin{table}[!t]
    \centering
    \caption{Distribution of comments across code entities and comment forms.}
    \label{tab:rq1}
    \resizebox{.98\textwidth}{!}{
        \input{tables/rq1_table}
    }

\end{table}


To answer RQ1, we calculate the frequency and proportion of CCPs for each pair of the code entity and comment form. In each cell of Table~\ref{tab:rq1}, the percentage following the frequency value shows the proportion relative to the total number of CCPs (i.e., 9,677). From the table, we observed that functions dominated code entities with 7,035 CCPs, while modules and classes were less frequently documented, accounting for 18.1\% and 9.2\%, respectively. This implies the preference for commenting on local and fine structures in quantum SDKs. In terms of comment forms, docstrings and block comments demonstrate similar proportions, whereas only a few CCPs (4.6\%) are attributed to inline comments. This observation suggests that Qiskit developers favor structured and systematic documentation like block comments and docstrings over ad-hoc remarks in the form of inline comments.  


More specifically, the comment preference of classes and modules deviates markedly from the overall pattern discussed above (i.e., close proportions of docstrings and block comments in total). For classes, the proportion of docstrings is noticeably greater than that of block comments, i.e., $7.7\% > 1.3\%$. The predominance of docstrings aligns with Python conventions, which recommend documenting classes using docstrings. In contrast, modules demonstrate an opposite relation to classes regarding frequencies for the two comment forms. The block comments involving 1,104 CCPs are the most frequent, followed by docstrings (610 CCPs), whereas inline comments are still rare, with only 38 CCPs. This may result from the fact that a module typically contains at most one module docstring, whereas one module may use multiple block comments for separate imports, global constants, and helper sections, leading to a greater number of block comments than docstrings.

\AnswertoRQ{\RQDistributionIndex{1}}{In terms of code entities, the majority of code comments (72.7\% of total CCPs) are concentrated at the function level. Regarding comment forms, the two most frequently used forms are docstrings and block comments, accounting for 47.9\% and 47.5\% of the total CCPs, respectively. By contrast, inline comments are considerably less common, representing only 4.6\% of the total CCPs.}




\subsubsection{\RQDistributionSect{2}}

\begin{table}[!t]
    \centering
    \small
    \caption{Comment density across different code entities.}
    \label{tab:rq2}
    \resizebox{.98\textwidth}{!}{
        \input{tables/rq2_table}
    }
\end{table}

We investigate the effort developers devote to writing comments for different code entities, as compared to the effort for entire code snippets. To quantify this, we adopt the comment density, a metric defined by the number of comment lines divided by the number of lines of code of the same code body~\cite{arafat2009comment, fenton2014software}. Besides, by analyzing this metric across code entities, we can identify the specific entities of the code where developers tend to concentrate their efforts and where such efforts are relatively lacking.

Table~\ref{tab:rq2} demonstrates the results by entity type. Generally, the comment density is 0.21, slightly above the average comment density of 0.19 reported in a previous work based on more than 5,000 open-source projects within the scope of CSE~\cite{arafat2009comment}. This result implies that the need for comments in QSE is no less important than that in CSE, and developers may even make more effort to write comments for quantum SDKs. At the entity level, functions reach a density of 0.39, which is well above the average 0.21. In contrast, code densities for classes and modules are respectively 0.15 and 0.13, both of which fall below 0.21. The difference in comment densities among the three entities reveals that function comments gain the most attention during development, which is consistent with the findings of RQ1 regarding the number of CCPs. In comparison, less effort is devoted to writing comments for classes and modules. 
Also, we observed that modules exhibited the lowest comment densities, despite having 21,420 comment lines more than those for classes. One reason could be that some module comments are misattributed to function comments, as the Python parsing tool LibCST sometimes struggles to distinguish between comments of the two levels.



\AnswertoRQ{\RQDistributionIndex{2}}{The overall comment density of 0.21, slightly higher than the reference 0.19, suggests that developers put at least as much effort into commenting quantum programs as they do for classical programs. Among the three entities, Qiskit developers have devoted the greatest effort to writing function comments relative to the size of entire code snippets.}

\subsubsection{\RQDistributionSect{3}}


\begin{figure}[!t]
    \centering

    \begin{minipage}{0.65\linewidth}
        \centering
        \includegraphics[width=0.5\textwidth]{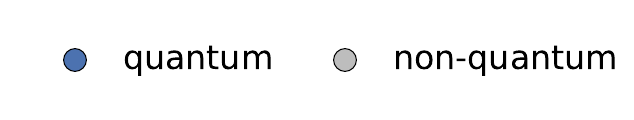}
        \label{fig:rq3_legend}
    \end{minipage}

    \begin{minipage}{0.52\linewidth}
        \centering
        \includegraphics[width=0.9\textwidth]{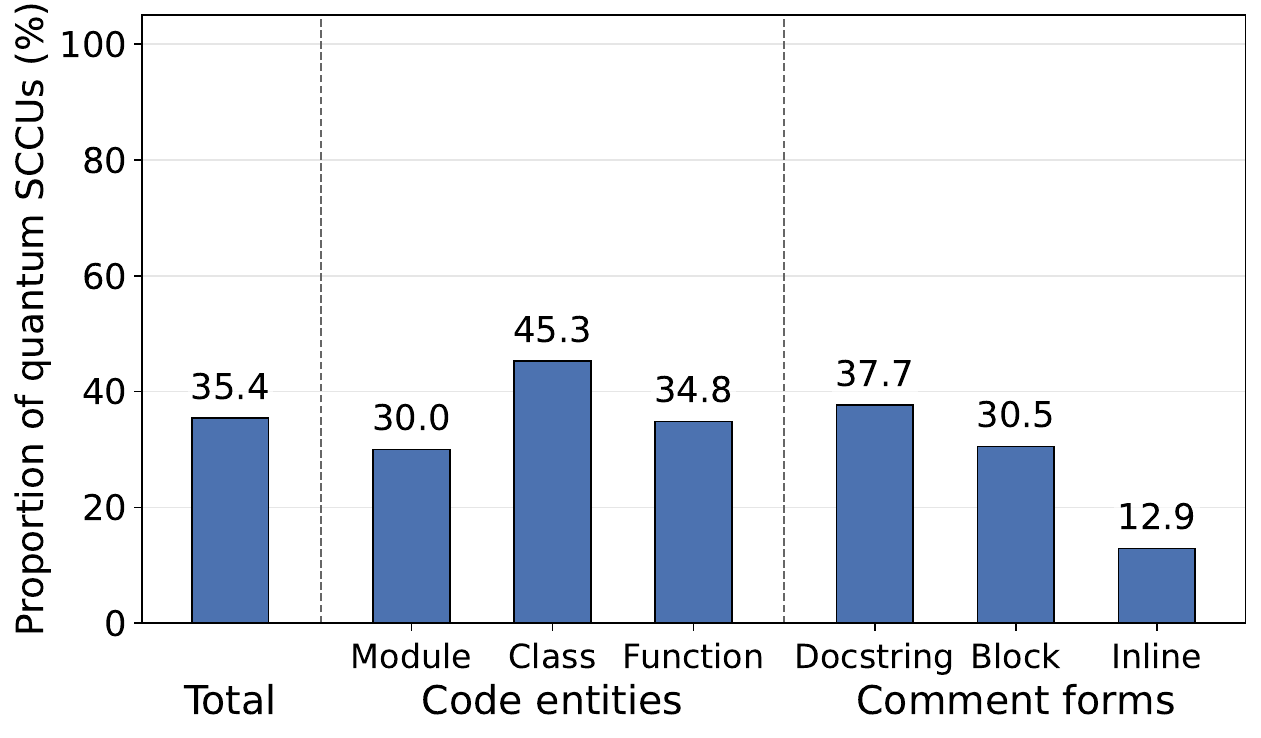}
        \subcaption{Proportion of \Quantum SCCUs over the total SCCUs in terms of one code entity or comment form}
        \label{fig:rq3_bar}
    \end{minipage}
    \hfill
    \begin{minipage}{0.45\linewidth}
        \centering
        \includegraphics[width=0.9\textwidth]{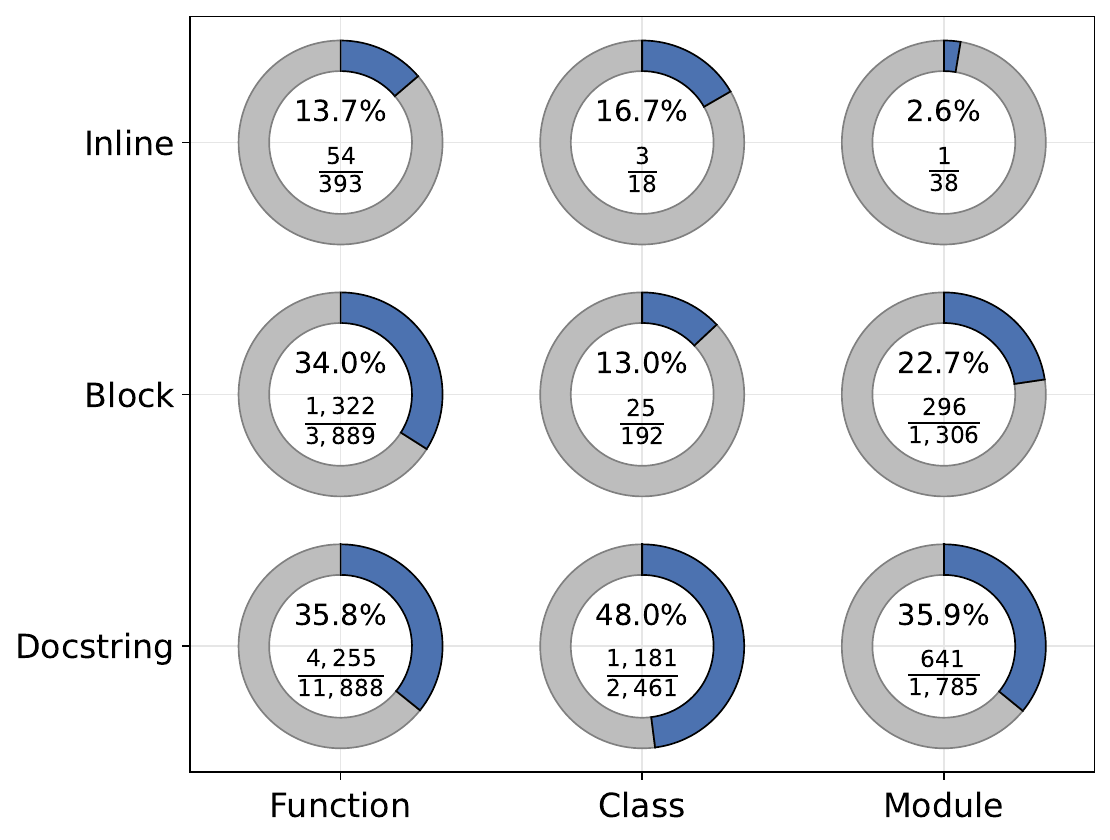}
        \subcaption{Proportion of \Quantum SCCUs over all SCCUs given a pair of code entity and comment form}
        \label{fig:rq3_donut}
    \end{minipage}

    \caption{SCCUs labeled as \Quantum or \NotQuantum SCCUs across code entities and comment forms.}
    \label{fig:rq3}

\end{figure}

Figure~\ref{fig:rq3_bar} illustrates the proportion of total \Quantum SCCUs upon one code entity or one comment form, where the bar above ``Total'' stands for the proportion of all \Quantum SCCUs within CC4Q. We observed that 35.4\% of SCCUs were labeled as \Quantum. This indicates that a moderate portion, but not all, of the comments in quantum SDKs are relevant to domain-specific knowledge. 

In terms of code entities in Figure~\ref{fig:rq3_bar}, \Quantum SCCUs in the class level exhibit the highest proportion (i.e., 45.3\%), while module-level \Quantum SCCUs constitute only 30.0\% of all module SCCUs, representing the lowest proportion. This reveals that comments at the module level are less frequently used to convey domain knowledge than the other two entities. With respect to comment forms, docstrings account for the highest proportion of \Quantum SCCUs (i.e., 37.7\%), whereas these \Quantum proportions for block and inline comments are both below the 35.4\% for ``Total''. Especially, only 12.9\% SCCUs in the form of inline comments show specific relevance to quantum topics. This may be because explaining quantum-specific knowledge easily understood usually requires formal and elaborate textual descriptions, whereas inline comments, being casual and short, often fail to achieve the supposed clarity. 

For specific combinations of entities and forms in Figure~\ref{fig:rq3_donut}, 4,255 SCCUs (19.4\% of the total 21,970 SCCUs in CC4Q) involve quantum-specific knowledge in function-level docstrings, representing a notably dominant proportion that aligns with the significant preference for writing function-level comments observed in RQ1. 
Within class-level docstrings, there are 1,181 \Quantum SCCUs with the proportion 48.0\%, suggesting that Qiskit developers are more likely to convey the domain-specific knowledge while writing docstrings for classes. In contrast, module-level inline comments contain only one \Quantum SCCU (2.6\%), the lowest in both count and proportion, indicating the negligible number of module-level inline comments conveying contents specific to quantum.

\AnswertoRQ{\RQDistributionIndex{3}} {The proportion of \Quantum SCCUs in CC4Q is 35.4\%, where 19.4\% of total SCCUs are \Quantum docstrings at the function level. Among all combinations of code entities and comment forms, the highest concentration of \Quantum SCCUs occurs in class-level docstrings, where they constitute 48.0\%.}

%% file: tables/rq1_table.tex



\begin{tabular}{
    >{\centering\arraybackslash}p{0.2\textwidth}|
    >{\centering\arraybackslash}p{0.2\textwidth}
    >{\centering\arraybackslash}p{0.2\textwidth}
    >{\centering\arraybackslash}p{0.2\textwidth}
    >{\centering\arraybackslash}p{0.2\textwidth}
}
    \toprule[1pt]
    \textbf{} & \textbf{Docstrings} & \textbf{Block comments} & \textbf{Inline comments} & \textbf{Total for forms} \\
    \midrule
    \textbf{Functions} & 3,283 (33.9\%) & 3,365 (34.8\%) & 387 (4.0\%) & 7,035 (72.7\%) \\
    
    \textbf{Classes} & 744 (7.7\%) & 130 (1.3\%) & 16 (0.2\%) & 890 (9.2\%) \\
    
    \textbf{Modules} & 610 (6.3\%) & 1,104 (11.4\%) & 38 (0.4\%) & 1,752 (18.1\%) \\
    
    \textbf{Total for entities} & 4,637 (47.9\%) & 4,599 (47.5\%) & 441 (4.6\%) & 9,677 (100.0\%) \\
    \bottomrule[1pt]
\end{tabular}

%% file: tables/rq2_table.tex

\begin{tabular}{
    >{\centering\arraybackslash}p{0.24\textwidth}|
    >{\centering\arraybackslash}p{0.24\textwidth}|
    >{\centering\arraybackslash}p{0.24\textwidth}|
    >{\centering\arraybackslash}p{0.24\textwidth}
}
    \toprule[1pt]
    \textbf{Code entities} & \textbf{Lines of comments} & \textbf{Lines of code snippets} & \textbf{Comment densities} \\
    \midrule
    Functions & 37,170 & 94,111 & 0.39 \\
    Classes & 12,635 & 86,577 & 0.15 \\
    Modules & 21,420 & 159,240 & 0.13 \\
    \midrule
    \textbf{Total} & 71,225 & 339,928 & 0.21 \\
    \toprule[1pt]
\end{tabular}
 

%% file: sections/empirical/developer-intent.tex
\subsection{Empirical Results and Analysis for the Developer-intent Perspective}

Understanding developers' intentions in code comments provides valuable insights for improving comment quality and code comprehension, particularly for quantum software strongly involving domain-specific purposes for development and maintenance. In \RQDeveloperIndex{1}, we explore differences in code comments between classical Java programs and Qiskit Python programs, as well as between \Quantum and \NotQuantum categories within Qiskit itself, in terms of involved developers' intentions. In \RQDeveloperIndex{2}, we perform a deeper analysis of the \DevOthers category---comments that do not clearly fit five developer-intent categories adopted in previous studies---to explore how quantum SDKs extend conventional ways of conveying intentions in comments. In this manner, we can characterize intent patterns and reveal domain-specific commenting behaviors in QSE.

\begin{figure}[!t]
    \centering
    \begin{subfigure}[t]{0.49\textwidth}
        \centering
        \includegraphics[width=0.9\textwidth]{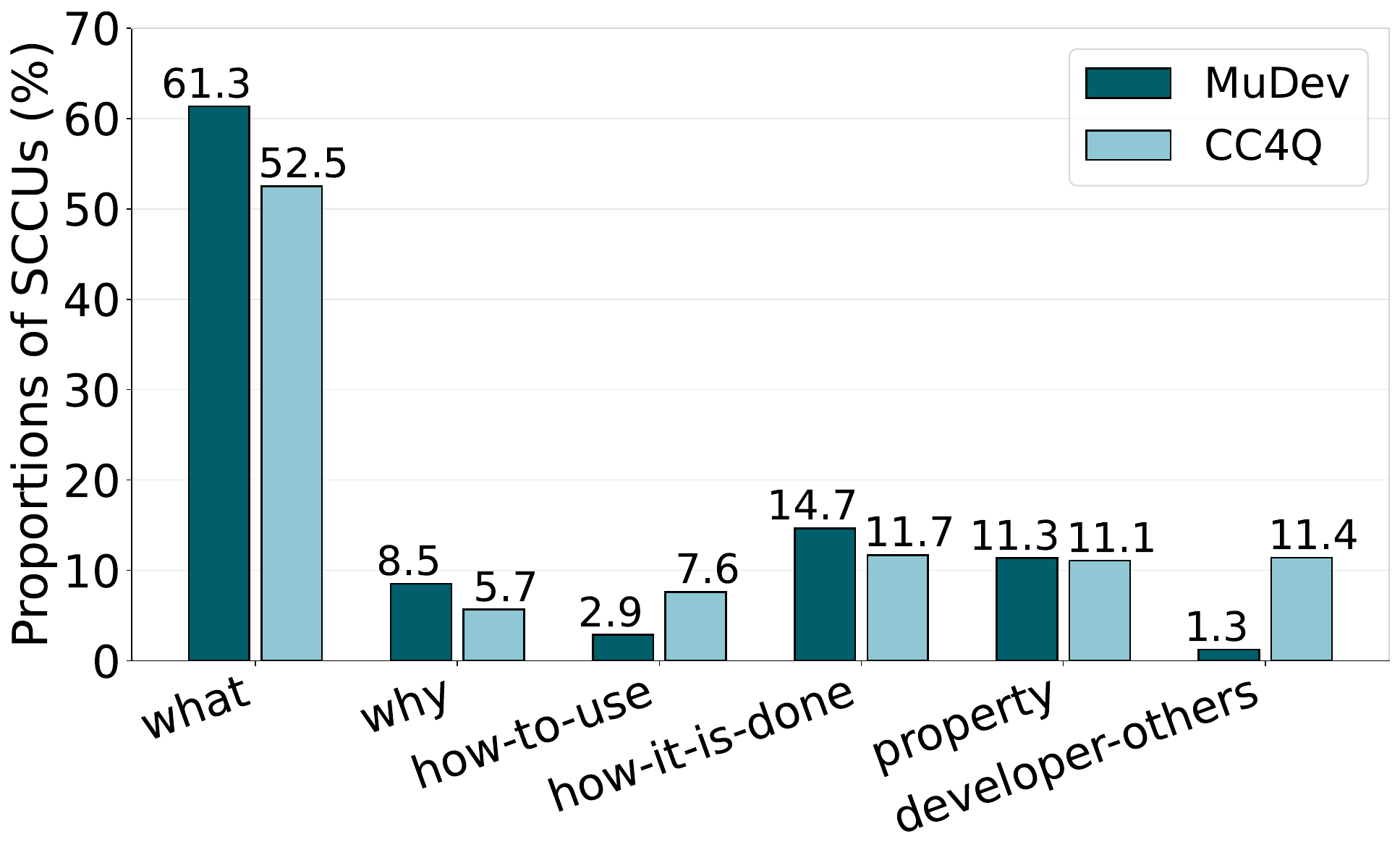}

        \caption{Comparison between CC4Q for QSE and MuDev for CSE}
        \label{fig:rq4-1}
    \end{subfigure}
    \hfill
    \begin{subfigure}[t]{0.49\textwidth}
        \centering
        \includegraphics[width=0.9\textwidth]{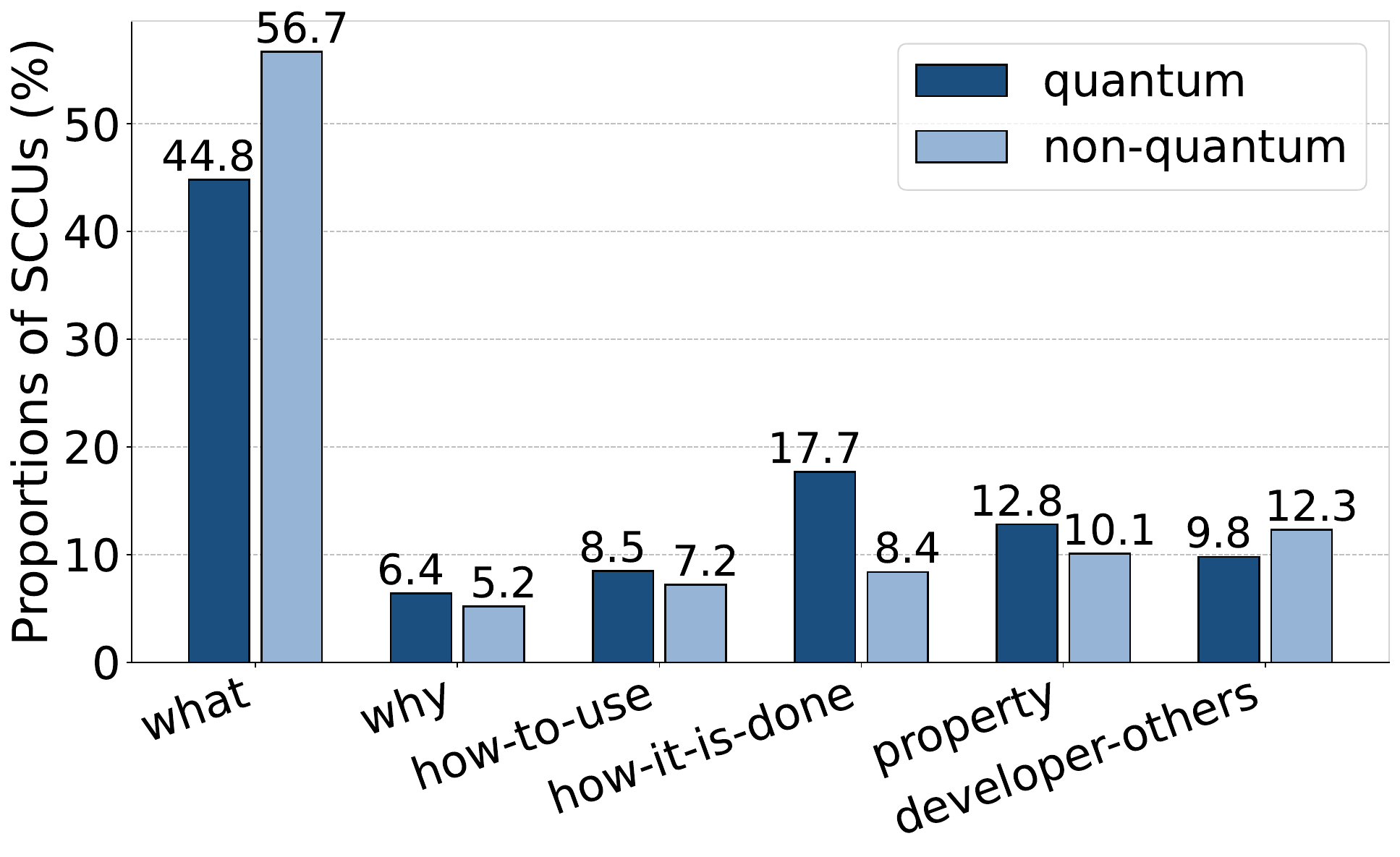}

        \caption{Comparison between \Quantum and \NotQuantum categories}
        \label{fig:rq4-2}
    \end{subfigure}
    \caption{Comparison of SCCUs in terms of developers' intentions}
\end{figure}

\subsubsection{\RQDeveloperSect{1}} 

To explore how developers’ intentions in code comments vary between QSE and CSE, we analyze the intent distribution of code comments using the taxonomy with six categories proposed in Section~\ref{sec:developer-intent}. Our analysis involves two levels of comparison for developers' intentions indicated in code comments: comparing Qiskit SDKs against classical Java programs (\textbf{COMP1}), and examining differences within Qiskit itself between \Quantum and \NotQuantum categories (\textbf{COMP2}). We reuse the data available in Mu et al.~\cite{mu2023developer} (noted as MuDev in our paper) in \textbf{COMP1}, where 20,000 CCPs from two Java datasets are manually classified into one of the first five developer-intent categories discussed in Section~\ref{sec:developer-intent} and the \Category{others} category\footnote{In our paper, we unify \Category{others} in MuDev as \DevOthers for convenience.}.


For \textbf{COMP1}, as shown in Figure \ref{fig:rq4-1}, we observed several distinctive patterns. Comments of \DevWhat category dominates across two datasets, where the proportions are 52.5\% for CC4Q and 61.3\% for MuDev. This demonstrates that developers universally prioritize descriptive explanations of code functionality. However, code comments in CC4Q exhibit notable differences in some other categories from those in MuDev. For instance, CC4Q possesses a higher proportion of \DevUse comments (7.6\% for CC4Q vs. 2.9\% for MuDev), reflecting the instructional commenting practices in quantum software, where Qiskit developers are inclined to provide extensive usage guidance for complex quantum operations. Also, a substantially higher proportion of \DevOthers comments are manifested in CC4Q (11.4\% for CC4Q vs. 1.3\% for MuDev). This indicates the existing developer-intent categories, along with our redefined \DevOthers, may be insufficient to fully capture the characteristics of code comments in quantum SDKs, where unique types of developer intentions may remain undiscovered.
 

Regarding \textbf{COMP2}, as illustrated in Figure~\ref{fig:rq4-2}, \Quantum comments exhibit distinctly different intent distributions compared to \NotQuantum comments. Notably, \Quantum comments show a lower proportion of \DevWhat category (44.8\% and 56.7\% for \Quantum and \NotQuantum, respectively), while demonstrating higher proportions in \DevDone (17.7\% for \Quantum vs. 8.4\% for \NotQuantum) and \DevProperty (12.8\% for \Quantum vs. 10.1\% for \NotQuantum). This pattern reveals that comments associated with quantum-specific knowledge require more comprehensive guidance beyond simple functional descriptions. Both types of \DevDone and \DevProperty comments, with their detailed information, can help practitioners to understand unfamiliar and complex quantum computing principles that should be involved in quantum software development and maintenance.



\AnswertoRQ{\RQDeveloperIndex{1}}{For both QSE and CSE, \DevWhat comments account for more than half of the developer-intent categories. The quantum SDK contains more \DevOthers comments than its classical counterpart. Compared to \NotQuantum comments, \Quantum comments are more frequently associated with the intentions of \DevProperty and \DevDone.}

\begin{figure}[!t]
    \centering
    \includegraphics[width=0.98\linewidth]{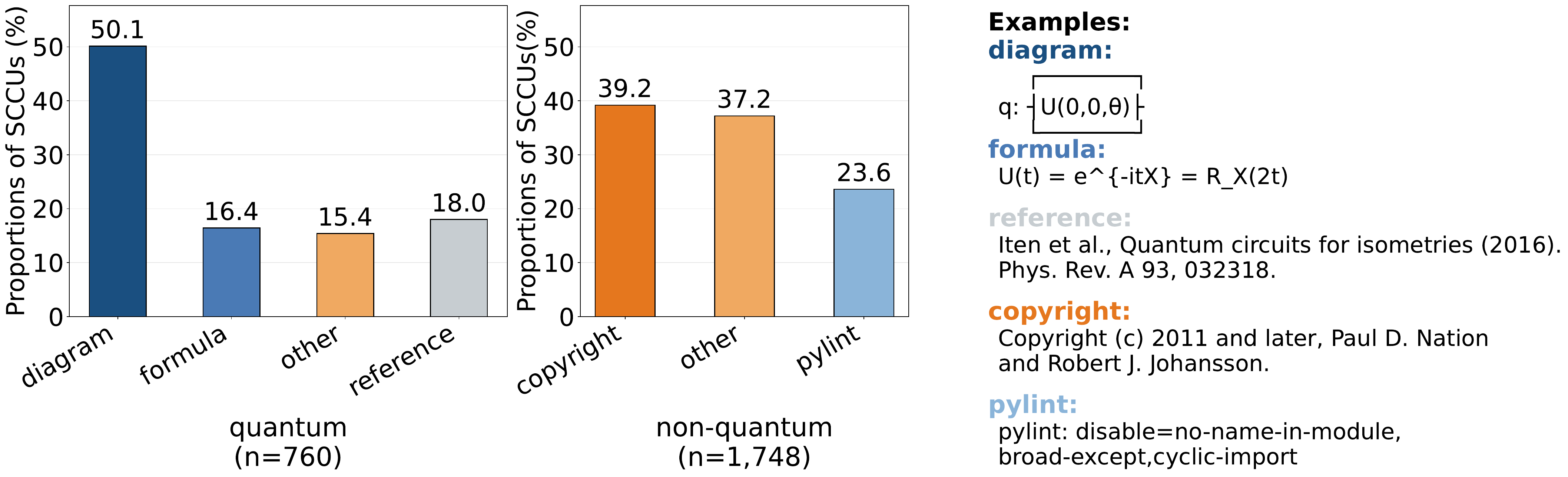}
    \caption{Results of extending \QuantumOthers to several patterns}
    \label{fig:rq5}
\end{figure}

\subsubsection{\RQDeveloperSect{2}}

Motivated by the results from \RQDeveloperIndex{1}, we further conduct a case study to answer \RQDeveloperIndex{2}, and explore the specific patterns of comments within the \QuantumOthers category to understand the novel practices in documenting Qiskit code. We adopt a semi-automatic approach to identify patterns based on the comment texts of \DevOthers SCCUs. We first use symbol matching to generate preliminary patterns from given candidates (e.g., \InlineCode{\textbackslash rangle} relevant to the Ket notation $\ket{\;}$ for mathematical formulas), and then human annotators review the outputs and correct the patterns misclassified.

As shown in Figure \ref{fig:rq5}, we discovered three distinct patterns involving 84.5\% of \Quantum SCCUs labeled as \DevOthers: visual circuit diagrams using ASCII art (50.1\% for \Category{diagram}), mathematical formulations describing quantum operations (16.4\% for \Category{formula}), and citations to academic papers (18.0\% for \Category{reference}). The three patterns reflect intentions of conveying complex quantum computing principles, which cannot be adequately expressed through traditional textual descriptions alone. Considering the three patterns' roles, for instance, circuit diagrams provide intuitive visual representations of quantum gate sequences; formulas present rigor and accurate mathematical operations in quantum information theory; and academic references bring educational values to the community. Furthermore, we identified two patterns within \NotQuantum SCCUs labeled as \DevOthers, with a total of 62.8\% corresponding to the two patterns.
The \NotQuantum comments are dominated by Copyright notices (39.2\% for \Category{copyright}) and code formatting directives like Pylint configurations (23.6\% for \Category{pylint}). This demonstrates valuable software engineering practices in the Python ecosystem, where developers rely on Pylint for code quality assurance and include copyright notices to address intellectual property rights in open-source projects.


\AnswertoRQ{\RQDeveloperIndex{2}}{Within the \DevOthers category, three patterns (i.e., \Category{diagram}, \Category{formula}, and \Category{reference}) are extracted from \Quantum comments and two patterns (i.e., \Category{copyright}, \Category{pylint}) are identified in \NotQuantum. Their relatively high proportions of comments including these patterns (i.e., 84.5\% for \Quantum and 62.8\% for \NotQuantum in total) highlight the importance of adopting these novel documentation practices in quantum SDKs.}


%% file: sections/empirical/quantum-specific.tex
\subsection{Empirical Results and Analysis for the Quantum-specific Perspective}
Considering that details for particular quantum topics have not been explored in the prior RQs, we focus on \Quantum code comments in this perspective to investigate what quantum-specific knowledge is conveyed through code comments in quantum SDKs. We first report the distribution of code comments in the six quantum-specific categories newly proposed in our paper to answer \RQQuantumIndex{1}. Extended from the developer-intent perspective discussed before, we compare quantum-specific and developer-intent taxonomies in \RQQuantumIndex{2} to figure out whether particular developers' intentions are more likely to be associated with certain quantum topics. Apart from that, we also perform an internal comparison among the six quantum-specific categories in \RQQuantumIndex{3}. More specifically, we quantify the semantic differences between quantum-specific categories to analyze their interrelations as well as the rationale for our proposed taxonomy. 

\subsubsection{\RQQuantumSect{1}}
\begin{figure}[!t]
    \centering
    \includegraphics[width=0.98\textwidth]{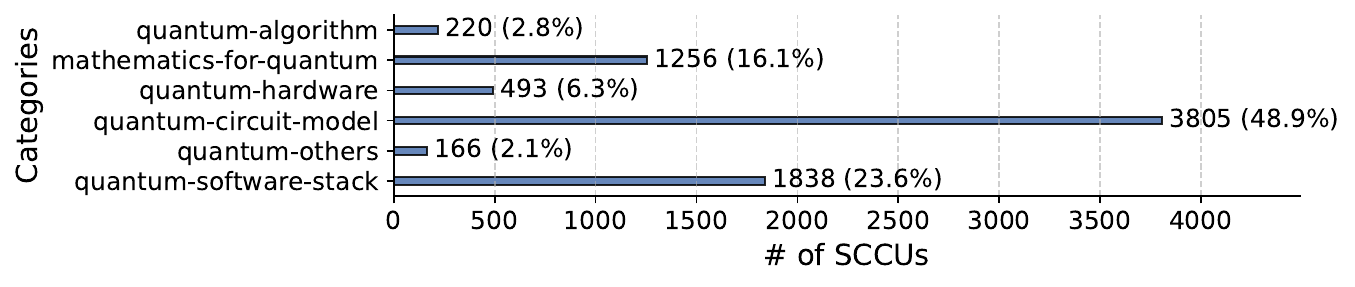}
    \caption{Distribution of SCCUs regarding the quantum-specific taxonomy}
    \label{fig:rq8}
\end{figure}
In this RQ, we display the distribution of SCCUs across six quantum-specific categories in Figure~\ref{fig:rq8}. There are 3,805 SCCUs classified as \QuantumCircuit, dominating the quantum-specific categories. This is because quantum circuit design and optimization play a leading part in software development for gate-based quantum systems. As a result, code comments are more likely to describe operations and procedures involved in the quantum circuit models. Additionally, 23.6\% and 16.1\% SCCUs respectively serve for the software stack and the related mathematics, both of which are critical for the design and implementation of quantum programs. For example, code comments labeled as \QuantumSoftware tell how to run quantum circuits with the backend, while \QuantumMath comments can depict the specification of a quantum subroutine due to the unitarity of quantum gates. We also observed that a small number of SCCUs (i.e., 220) corresponded to the entire quantum algorithms. On the one hand, current code comments tend to present implementation details of quantum algorithms, such as gate placement, thereby these code comments being classified as \QuantumCircuit. On the other hand, according to the view in~\cite{shor2003haven}, there exists a limited number of applicable quantum algorithms, so the volume of code comments pertaining to specific quantum algorithms remains small, correspondingly.  

\AnswertoRQ{\RQQuantumIndex{1}}{About half of code comments (48.9\%) are directly associated with the quantum circuit model. Also, a moderate number of code comments correspond to the quantum software stack (23.6\%) and the mathematics for quantum-based terms (16.1\%). Meanwhile, quantum hardware (6.3\%) and quantum algorithms (2.8\%) receive relatively little attention in the code comments.}

\subsubsection{\RQQuantumSect{2}}
\begin{figure}[!t]
    \centering
    \begin{subfigure}[t]{0.49\textwidth}
        \centering
        \includegraphics[width=0.98\textwidth]{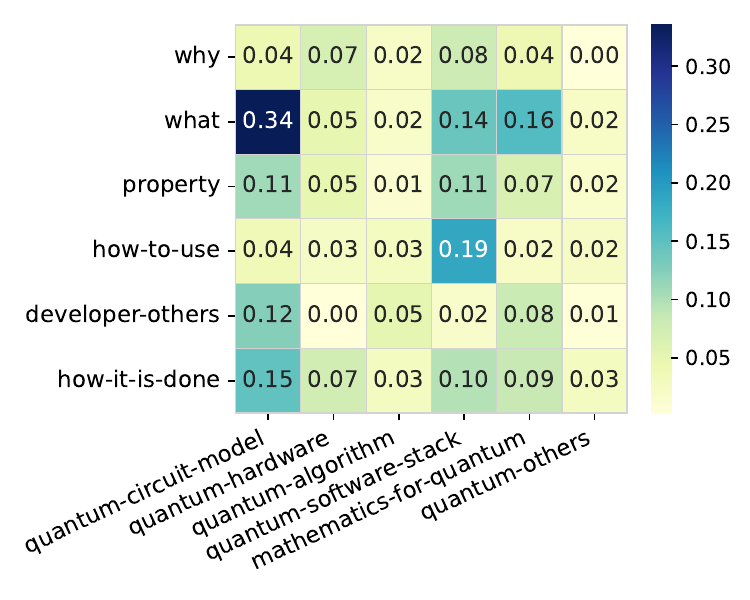}
        \caption{Heatmap for Jaccard similarity}
        \label{rq9:Jaccard}
    \end{subfigure}
    \begin{subfigure}[t]{0.49\textwidth}
        \centering
        \includegraphics[width=0.98\textwidth]{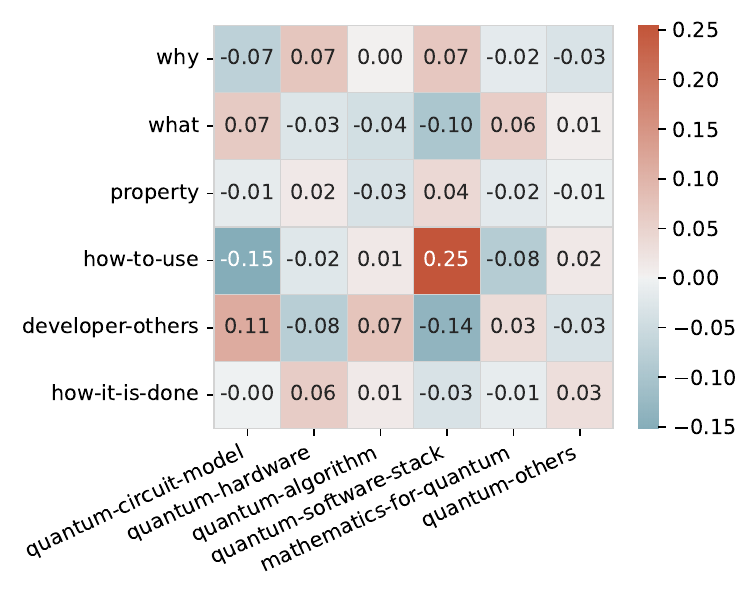}
        \caption{Heatmap for Pearson coefficient}
        \label{rq9:Pearson}
    \end{subfigure}
    \caption{Relationship between the two groups of developer-intent categories and quantum-specific categories}
\end{figure}
This RQ measures the relationship between developer-intent and quantum-specific taxonomies in terms of co-occurrence and correlation. We employ Jaccard similarity~\cite{jaccard1912distribution} to assess the co-occurrence of developer-intent and quantum-specific categories in one SCCU. This metric depends on the size of the intersection and union sets relevant to the two kinds of categories, where a high similarity indicates that the two categories are likely to co-occur. The Pearson coefficient~\cite{pearson1895vii} is used to quantify the statistical correlation between the above two kinds of categories. Statistically, this metric can further capture the positive and negative associations of two categories.

As shown in Figure~\ref{rq9:Jaccard}, most of the category pairs manifest low similarity with values below 0.10. This indicates that the developers' intentions on a \Quantum code comment should be diverse. Still, the similarity between \DevWhat and \QuantumCircuit is the greatest (i.e., 0.34) among those of all category pairs, revealing the importance of comments to describe the functionalities of quantum circuits. Comments of \DevUse and \QuantumSoftware are coupled with a similarity of 0.19, i.e., the second greatest value. Taking an example to support this observation, many comments display code examples and usually aim at teaching developers to use the APIs provided in the software stack appropriately.

Regarding the correlation in Figure~\ref{rq9:Pearson}, we noticed that all the category couples except \DevUse and \QuantumSoftware had the absolute value of the Pearson coefficient below 0.2. This implies that the developer-intent and quantum-specific taxonomies are statistically independent of each other, and they present almost different perspectives of the \Quantum comments. Furthermore, this result suggests that \DevUse and \QuantumSoftware are likely to appear in the same SCCU to some extent, once code comments involve the two categories. Many comments that present code examples to elaborate on the usage of APIs could relate to this phenomenon. In comparison, the couple of \DevWhat and \QuantumCircuit yielding a Pearson coefficient with a small absolute value (i.e., $0.07$) reflects weak linear correlation, even though the two categories exhibit a moderate degree of co-occurrence. This could be due to the high proportions of \DevWhat and \QuantumCircuit comments. Large sample sizes may result in a moderate Jaccard similarity, as this metric focuses on the size of the sets, yet may not necessarily imply a direct linear relationship.

\AnswertoRQ{\RQQuantumIndex{2}}{Most pairs of developer-intent and quantum-specific categories show a relatively weak relationship in terms of co-occurrence and correlation. However, \DevWhat and \QuantumCircuit tend to co-occur. Two categories \DevUse and \QuantumSoftware demonstrate a slight degree of co-occurrence and even exhibit a moderate linear correlation.}

\subsubsection{\RQQuantumSect{3}}
\begin{figure}[!t]
    \centering
    \begin{subfigure}[t]{0.49\textwidth}
        \centering
        \includegraphics[width=0.98\textwidth]{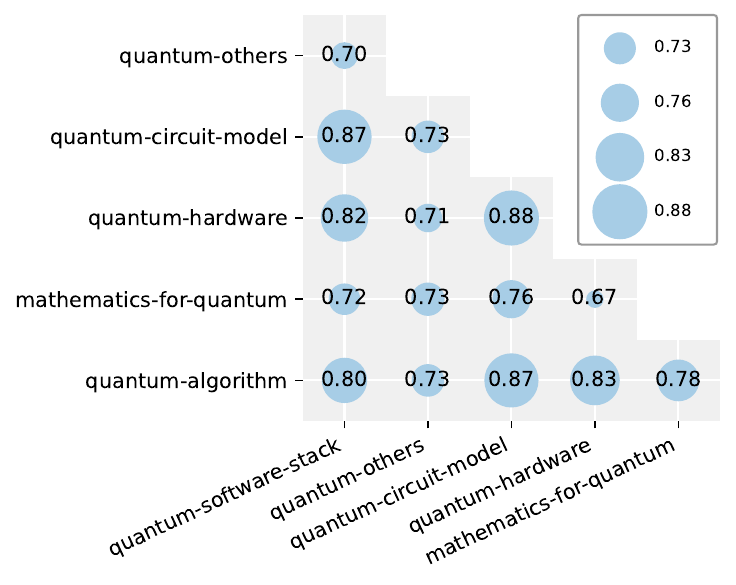}
        \caption{Bubble chart for cosine similarity}
        \label{rq10:bubble}
    \end{subfigure}
    \begin{subfigure}[t]{0.49\textwidth}
        \centering
        \includegraphics[width=0.98\textwidth]{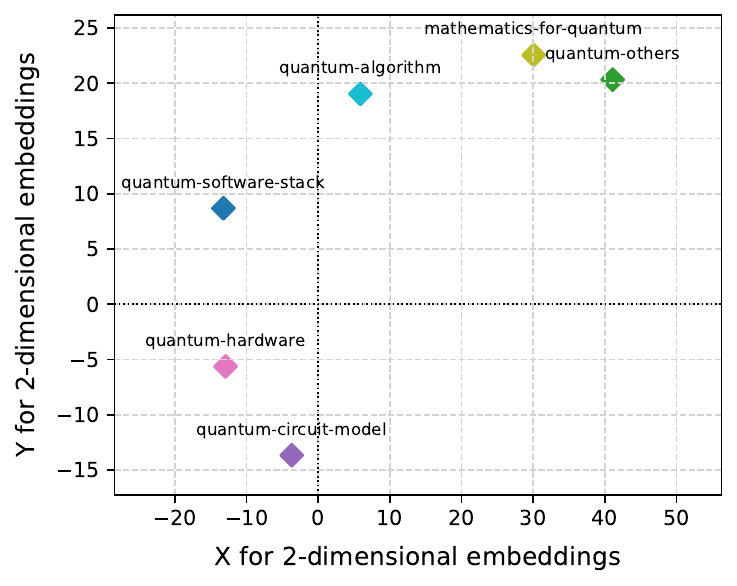}
        \caption{Two-dimension centroids with coordinates $(X, Y)$}
        \label{rq10:centroid}
    \end{subfigure}
    \caption{Semantic relationship among quantum-specific categories}
\end{figure}

In this RQ, we aim to investigate the semantic differences among six quantum-specific categories in terms of SCCU embeddings. We first employ Sentence Transformers in Hugging Face~\cite{reimers2019sentence}, a Python framework for state-of-the-art sentence and text embeddings, to vectorize each \Quantum SCCU. Then, the centroid of each category is calculated as the average of the embeddings of all the included SCCUs. Next, the cosine similarity~\cite{schutze2008introduction} widely used in textual analysis is adopted to quantify the semantic differences via the centroids corresponding to the two categories. For better visualization, we utilize t-SNE (t-distributed Stochastic Neighbor Embedding)~\cite{maaten2008visualizing} to project the high-dimensional embedding into a two-dimensional plane, where t-SNE is capable of retaining the local proximity relationships during dimension reduction.

According to Figures~\ref{rq10:bubble} and~\ref{rq10:centroid}, each of the six categories is well distinguished from the others. Regarding the cosine similarity, the coefficients of all the category pairs range from 0.67 to 0.88. Since all categories pertain to quantum-based knowledge, their semantic distinctions remain within a limited scope. Concerning the upper bound of cosine similarity, this can support the rationality of our proposed taxonomy, as no redundant category is included. A redundant category would otherwise yield a cosine similarity of nearly 1.00 with at least one existing category. When observing the concrete similarity of specific category pairs, \QuantumCircuit is semantically close to \QuantumAlgorithm, \QuantumHardware, and \QuantumSoftware, since the corresponding cosine similarities are either 0.87 or 0.88. Intuitively, this holds true, as the quantum circuit model, for instance, can visualize specific quantum algorithms, support transpilation tailored to quantum hardware, and connect with the relevant data structures in software development. Besides, we found that any category couple including \QuantumMath or \QuantumOthers showed the similarity of no more than 0.78, demonstrating weak semantical overlap with the other categories. One reasonable explanation should be that \QuantumMath code comments involve many mathematical expressions instead of natural language, and \QuantumOthers category is proposed due to its difficulty in being classified into other categories. 

\AnswertoRQ{\RQQuantumIndex{3}}{With the similarity from 0.67 to 0.88, the categories within the quantum-specific taxonomy are semantically distinguishable but share the quantum topic. The four categories (\QuantumCircuit, \QuantumAlgorithm, \QuantumHardware, and \QuantumSoftware) are semantically close, whereas \QuantumMath and \QuantumOthers are relatively distinct from other four categories.}

%% file: sections/sec5_threatsValidity.tex
\section{Threats to Validity}
\label{sec:threats}
The first threat of this empirical study is the reliability of labeled categories. Despite a careful and systematic annotating procedure adopted in our work, the subjectivity brought by human involvement inevitably introduces potential biases in the labels. Besides, following the existing works~\cite{chen2021my, leclair2019neural-funcom, mu2023developer, zhai2020cpc}, we use the machine learning models to infer the categories of numerous SCCUs, but the uncertainty resulting from these models impacts the labeling reliability as well. To this end, we employ BERT to mitigate this threat, a pre-trained deep learning model more advanced than some models previously adopted, such as Decision Tree, Random Forest, and Convolutional Neural Network. In the future, it will remain meaningful to present a clean and refined version of our CC4Q with additional effort taken.

Another threat to our findings is the scope of the dataset, i.e., only one Qiskit repository with Python code being mined. As the first dataset on code comments for quantum SDKs, we prioritize this repository according to Qiskit's popularity for quantum programming and the generic interfaces involved in this repository. Hence, our findings, based on the relatively large dataset, are promising to represent general use cases of quantum SDKs. Even so, these findings could not always be generalized to other contexts, such as the D-Wave Ocean SDK\footnote{https://www.dwavequantum.com/solutions-and-products/ocean/} for quantum annealers instead of gate-based quantum systems, and the quantum machine learning framework discussed in~\cite{zhao2023empirical}, which zooms into a specific branch at the application level. In addition, as most mainstream and advanced quantum SDKs use Python rather than Java as their host language~\cite{zhao2020quantum}, it is challenging to avoid latent bias when comparing Qiskit Python code with classical Java code in analyses of code comments between QSE and CSE.


%% file: sections/sec6_discussion.tex
\section{Discussion}
\label{sec:discussion}
From the three perspectives, we discuss our empirical findings and offer practical suggestions drawing on exemplary instances from the official Qiskit repository.
\subsection{Structure and Coverage of Code Comments for Quantum Programs}
In the Qiskit official repository, most code comments are written in the form of docstrings and block comments, suggesting that developers should attach importance to well-structured comments in their own quantum projects. Moreover, the observation that most of the comments in Qiskit are written for fine-grained entities like functions indicates the necessity for developers to document more specific implementation details, given the inherently esoteric principles behind quantum computing. Based on the average comment density of 0.21 reported in~\RQDistributionIndex{2}, we recommend that the volume of comments written by developers should scale with project size, and it will be better to ensure around 21\% code lines written for comments in scenarios requiring high maintainability. 
In~\RQDistributionIndex{3}, we observe that comments in quantum SDKs integrate established conventions from classical computing with domain-specific terminologies from quantum topics.
Especially, quantum-specific terminologies are suggested to be explained in docstrings that not only allow rich textual descriptions but also provide a clear overview of the documented entity.
Additionally, we note that inconsistent commenting conventions across programming languages may pose challenges for documenting quantum code in cross-language contexts.
For instance, inline comments are popular in developing classical Java programs~\cite{huang2023comparative}, whereas we observe that only a few inline comments appear in Qiskit Python programs. This is largely because our identification of inline comments strictly requires them to reside on the same line as the corresponding statement, in accordance with the PEP 8-style guide~\cite{van2001pep}. Consequently, we suggest that future work could explore standardized documentation practices suitable for different quantum SDKs built upon respective host languages.

\subsection{Inclination of Developers' Intentions in the Quantum SDK}
In view of developers' intentions conveyed in code comments, we find some consistencies between CSE and QSE. Apart from classical programs, for instance, offering definitions and summarizing the functionality of key subjects are still essential in quantum programs, on which more than half of developers' intentions focus. Still, more \DevUse comments in quantum SDKs than those of classical counterparts highlight that developers should offer details on how to use quantum APIs (e.g., via example code), since many current users lack proficiency in these quantum SDKs with continuous iterations. Based on the results of~\RQDeveloperIndex{1}, we suggest that, for the purpose of conveying quantum-specific knowledge, code comments should not be restricted to simple summaries but also elaborate on specific details, for example, how a code block is done and the properties of a method. \RQDeveloperIndex{2} extends the \DevOthers category based on the unique patterns reflected in Qiskit. Diagrams using ASCII art, mathematical formulas, academic references, program copyrights, and Pylint components are heavily engaged in Qiskit, yet they are not reported in the developer-intent taxonomy proposed for classical Java programs. This discrepancy emphasizes the need to explore documentation practices specifically tailored to QSE, ensuring that developers' intentions are conveyed to users with clarity and precision. We recommend that developers with expertise in quantum physics and quantum computing incorporate diagrams, formulas, and references in the code comments to improve not only the readability and maintainability of the code but also the educational value of quantum software projects. Also, we encourage future research work to systematically explore a broader range of developer-intent comment patterns in quantum SDKs, as only a small case study is attempted in \RQDeveloperIndex{2}.

\subsection{Quantum-specific Knowledge for Quantum Software Development}
The dominance of \QuantumCircuit in the quantum-specific taxonomy implies that documenting quantum software, particularly quantum subroutines, depends on developers' familiarity with the fundamentals of quantum computing. The frequent co-occurrence of \DevWhat and \QuantumCircuit indicates that summarizing quantum circuit functionality is essential, reflecting the demand for foundational knowledge. While the circuit model has received substantial attention in the QSE community, the moderate proportions of \QuantumHardware and \QuantumSoftware comments indicate that developing quantum software should extend beyond program design to incorporate systematic thinking, such as the properties of physical quantum hardware and the software aspects of system architecture.
Based on the close link between \DevUse and \QuantumSoftware revealed in \RQQuantumIndex{2}, we strongly encourage individual developers and researchers to increase their commenting effort on the usage of components in quantum software stacks, since documenting these applicable components (e.g., backends and compilers) can be beneficial for both academia and industry.
The \QuantumMath comments with the third greatest proportion in the quantum-specific taxonomy highlight the importance of mathematics in quantum software development, where mathematics provides a formal framework grounded in quantum information theory and quantum mechanics. 
However, the relatively weak semantic similarity between \QuantumMath and the other quantum-specific categories implies that developers should make considerable effort to learn mathematics when intending to cover quantum-related knowledge necessary for quantum software development.
Overall, our findings align with prior calls~\cite{pezze20252030} for educating quantum software engineers, which can foster quantum software development and maintenance through high-quality code comments.




%% file: sections/sec7_relatedwork.tex
\section{Related Work}
\label{sec:related}

\summary{Code documentation for CSE} Java has gained the most attention in CSE research on code documentation, due to its popularity among developers and educators~\cite{rai2022review}. Some prevailing datasets, such as Funcom~\cite{leclair2019recommendations} and TLC~\cite{hu2018summarizing-tlc}, are available for research on code documentation of Java projects.
These datasets have been widely used for tasks such as comment classification~\cite{mu2023developer}, summarization~\cite{SuMcMillan2024}, and comment quality studies~\cite{EvtikhievBogomolov2023, RoyFakhoury2021}, highlighting their important role in CSE code documentation research.
Differently, our paper presents the dataset for Python code, due to Python's dominance in developing applicable quantum programs. Recently, several works~\cite{zhai2020cpc, chen2021my, mu2023developer} have categorized Java code comments based on developers' intentions, and further investigated some issues like automatic classification and propagation of code comments.
However, unlike general-purpose software, comments in quantum programs often contain domain-specific concepts such as quantum gates, circuits, hardware constraints, or mathematical formalisms, which are not covered by existing CSE datasets. This motivates the need for a quantum-specific taxonomy and a dedicated dataset.
In our paper, we validate the developer-intent taxonomy for code comments in Qiskit. 
Owing to the absence of code comment datasets for quantum computing, our paper focuses on the empirical analysis of Qiskit code comments. The resulting dataset and findings are promising for evaluating techniques aimed at improving code comment quality, and supporting automatic comment classification and generation. 

\summary{Datasets for software quality within QSE} Several works have introduced datasets through empirical analysis of quantum SDKs to assess software quality. Analogous to our paper, these studies often adopted empirical approaches through mining and analyzing the open-source repositories relevant to quantum SDKs, such as Paltenghi et al.~\cite{paltenghi2022bugs} examining real-world bugs, Zhao et al.~\cite{zhao2021bugs4q, zhao2023bugs4q} collecting general Qiskit bug-fix pairs, and Zhang et al.~\cite{zhang2023identifying} identifying flaky tests related to official quantum SDKs. Despite these contributions, most of these empirical studies concentrate on post-execution quality assurance of software, but the steps taken by developers during the development and maintenance processes remain almost underexplored. This gap motivates our paper to aim at developers and investigate code comments as a means of quality assurance.

\summary{Code documentation for QSE} 
Several works have been conducted directly for the code documentation or relevant issues of the QSE.
For the work by d’Aloisio et al.~\cite{d2024exploring}, \textbf{Large Language Models (LLMs)} were leveraged to generate code summarization of quantum programs. Instead of detailed code comments, only the overall explanations for circuit-like quantum programs implementing quantum algorithms are produced. Hence, the scope of code documentation is limited to quantum algorithms, corresponding to the \QuantumCircuit or \QuantumAlgorithm categories. Meanwhile, fine-grained documentation, like inline comments, block comments, and docstrings, have seldom been involved.
In addition, Yu et al.~\cite{YuLi2025} proposed QuAInth, an LLM-based approach that generates comments for application-oriented quantum programs. This approach produces high-level descriptions rather than fine-grained developer comments and is limited to application-oriented quantum programs. Hence, this motivates our work to explore diverse structured code comments and abundant programs from the perspective of quantum SDKs. Apart from the generation-based studies that particularly leveraged LLMs, Ishimoto et al.~\cite{IshimotoNakamura2024} concentrated on a large-scale manual analysis of \textbf{Self-Admitted Technical Debt (SATD)} comments in quantum projects. This study revealed quantum-specific SATD patterns, but covered only SATD-related comments rather than general code documentation. These existing studies strengthen the importance of code comments for QSE, and even the two studies~\cite{d2024exploring, YuLi2025} moved forward techniques on code comment generation. However, no existing work has examined developer-written comments in quantum SDKs or provided a systematic dataset for them, where this issue is reflected in the need for human involvement to evaluate the generation quality~\cite{d2024exploring, YuLi2025}. Therefore, we zoom into the Qiskit official repository to present a more comprehensive view of code comments involved in the quantum software development.

%% file: sections/sec8_conclusion.tex
\section{Conclusion and Future Work}
\label{sec:conclusion}
Our paper presents CC4Q, the first dataset of code comments for quantum \textbf{Software Development Kits (SDKs)}, intended to support and extend research on code documentation in \textbf{Quantum Software Engineering (QSE)}. Considering the non-intuitive characteristics of quantum physics and quantum computing, we provide developers with a new taxonomy to outline the quantum-based knowledge conveyed in the code comments of the Qiskit official repository. Upon our constructed CC4Q, a comprehensive empirical study is conducted from the structure-based, developer-intent, and quantum-specific perspectives. The empirical results can reveal exemplary paradigms of writing high-quality code comments, and also outline future directions to facilitate the maintainability of quantum programs via code comments. Future studies may aim to enlarge the dataset scalability or provide a cleaned version. CC4Q is also expected to contribute to the evaluation of relevant automated techniques, such as code comment generation for quantum programs.

%% file: sections/sec9_dataCodeAvailability.tex
\section*{Data and Code Availability}
\addcontentsline{toc}{section}{Data and Code Availability}
We have the artifact of our paper available at~\cite{cc4q2025artifact}, where the relevant code, data, and documentation are included.
 

%% file: sections/sec10_acknowledgments.tex
\section*{Acknowledgments}

Yuechen Li was partly supported by the China Scholarship Council (CSC) under No. 202506020099. This study was funded by the National Natural Science Foundation of China under Grant 62372021 and the National Key R\&D Program of China under Grant 2024YFB3311503.

%% file: sections/sec0_ref.bib
@inproceedings{yang2019survey,
  author = {Yang, Bai and Liping, Zhang and Fengrong, Zhao},
  booktitle = {Proceedings of the 2019 3rd International Conference on Management Engineering, Software Engineering and Service Sciences},
  pages = {45--51},
  title = {A Survey on Research of Code Comment},
  year = {2019}
}

@article{piattini2021toward,
  author = {Piattini, Mario and Serrano, Manuel and Perez-Castillo, Ricardo and Petersen, Guido and Hevia, Jose Luis},
  journal = {IT Professional},
  number = {1},
  pages = {62--66},
  publisher = {IEEE},
  title = {Toward a Quantum Software Engineering},
  volume = {23},
  year = {2021}
}

@article{ying2012floyd,
  author = {Ying, Mingsheng},
  journal = {ACM Transactions on Programming Languages and Systems (TOPLAS)},
  number = {6},
  pages = {1--49},
  publisher = {ACM New York, NY, USA},
  title = {Floyd--Hoare Logic for Quantum Programs},
  volume = {33},
  year = {2012}
}

@article{aleksandrowicz2019qiskit,
  author = {Aleksandrowicz, Gadi and Alexander, Thomas and Barkoutsos, Panagiotis and Bello, Luciano and Ben-Haim, Yael and Bucher, David and Cabrera-Hernández, Francisco Jose and Carballo-Franquis, Jorge and Chen, Adrian and Chen, Chun-Fu and Chow, Jerry M. and Córcoles-Gonzales, Antonio D. and Cross, Abigail J. and Cross, Andrew and Cruz-Benito, Juan and Culver, Chris and De La Puente González, Salvador and De La Torre, Enrique and Ding, Delton and Dumitrescu, Eugene and Duran, Ivan and Eendebak, Pieter and Everitt, Mark and Faro Sertage, Ismael and Frisch, Albert and Fuhrer, Andreas and Gambetta, Jay and Godoy Gago, Borja and Gomez-Mosquera, Juan and Greenberg, Donny and Hamamura, Ikko and Havlicek, Vojtech and Hellmers, Joe and Herok, tukasz and Horii, Hiroshi and Hu, Shaohan and Imamichi, Takashi and Itoko, Toshinari and Javadi-Abhari, Ali and Kanazawa, Naoki and Karazeev, Anton and Krsulich, Kevin and Liu, Peng and Luh, Yang and Maeng, Yunho and Marques, Manoel and Martín-Fernández, Francisco Jose and McClure, Douglas T. and McKay, David and Meesala, Srujan and Mezzacapo, Antonio and Moll, Nikolaj and Moreda Rodríguez, Diego and Nannicini, Giacomo and Nation, Paul and Ollitrault, Pauline and O'Riordan, Lee James and Paik, Hanhee and Pérez, Jesús and Phan, Anna and Pistoia, Marco and Prutyanov, Viktor and Reuter, Max and Rice, Julia and Rodríguez Davila, Abdón and Putra Rudy, Raymond Harry and Ryu, Mingi and Sathaye, Ninad and Schnabel, Chris and Schoute, Eddie and Setia, Kanav and Shi, Yunong and Silva, Adenilton and Siraichi, Yukio and Sivarajah, Seyon and Smolin, John A. and Soeken, Mathias and Takahashi, Hitomi and Tavernelli, Ivano and Taylor, Charles and Taylour, Pete and Trabing, Kenso and Treinish, Matthew and Turner, Wes and Vogt-Lee, Desiree and Vuillot, Christophe and Wildstrom, Jonathan A. and Wilson, Jessica and Winston, Erick and Wood, Christopher and Wood, Stephen and Wörner, Stefan and Akhalwaya, Ismail Yunus and Zoufal, Christa},
  journal = {},
  title = {Qiskit: An Open-source Framework for Quantum Computing},
  year = {2019}
}

@article{rai2022review,
  author = {Rai, Sawan and Belwal, Ramesh Chandra and Gupta, Atul},
  journal = {ACM Transactions on Intelligent Systems and Technology (TIST)},
  number = {5},
  pages = {1--44},
  publisher = {ACM New York, NY},
  title = {A Review on Source Code Documentation},
  volume = {13},
  year = {2022}
}

@inproceedings{zhang2023identifying,
  author = {Zhang, Lei and Radnejad, Mahsa and Miranskyy, Andriy},
  booktitle = {2023 ACM/IEEE International Symposium on Empirical Software Engineering and Measurement (ESEM)},
  organization = {IEEE},
  pages = {1--7},
  title = {Identifying Flakiness in Quantum Programs},
  year = {2023}
}

@inproceedings{zhao2021bugs4q,
  author = {Zhao, Pengzhan and Zhao, Jianjun and Miao, Zhongtao and Lan, Shuhan},
  booktitle = {2021 36th IEEE/ACM International Conference on Automated Software Engineering (ASE)},
  organization = {IEEE},
  pages = {1373--1376},
  title = {Bugs4q: A Benchmark of Real Bugs for Quantum Programs},
  year = {2021}
}

@inproceedings{zhao2023empirical,
  author = {Zhao, Pengzhan and Wu, Xiongfei and Luo, Junjie and Li, Zhuo and Zhao, Jianjun},
  booktitle = {2023 IEEE International Conference on Quantum Software (QSW)},
  organization = {IEEE},
  pages = {68--75},
  title = {An Empirical Study of Bugs in Quantum Machine Learning Frameworks},
  year = {2023}
}

@article{zhao2023bugs4q,
  author = {Zhao, Pengzhan and Miao, Zhongtao and Lan, Shuhan and Zhao, Jianjun},
  journal = {Journal of Systems and Software},
  pages = {111805},
  publisher = {Elsevier},
  title = {Bugs4q: A Benchmark of Existing Bugs to Enable Controlled Testing and Debugging Studies for Quantum Programs},
  volume = {205},
  year = {2023}
}

@article{paltenghi2022bugs,
  author = {Paltenghi, Matteo and Pradel, Michael},
  journal = {Proceedings of the ACM on Programming Languages},
  number = {OOPSLA1},
  pages = {1--27},
  publisher = {ACM New York, NY, USA},
  title = {Bugs in Quantum Computing Platforms: An Empirical Study},
  volume = {6},
  year = {2022}
}

@inproceedings{d2024exploring,
  author = {d'Aloisio, Giordano and Fortz, Sophie and Hanna, Carol and Fortunato, Daniel and Bensoussan, Avner and Mendiluze Usandizaga, E{\~n}aut and Sarro, Federica},
  booktitle = {Proceedings of the 18th ACM/IEEE International Symposium on Empirical Software Engineering and Measurement},
  pages = {475--481},
  title = {Exploring LLM-driven Explanations for Quantum Algorithms},
  year = {2024}
}

@inproceedings{leclair2019neural-funcom,
  author = {LeClair, Alexander and Jiang, Siyuan and McMillan, Collin},
  booktitle = {2019 IEEE/ACM 41st International Conference on Software Engineering (ICSE)},
  organization = {IEEE},
  pages = {795--806},
  title = {A Neural Model for Generating Natural Language Summaries of Program Subroutines},
  year = {2019}
}

@inproceedings{hu2018summarizing-tlc,
  author = {Hu, Xing and Li, Ge and Xia, Xin and Lo, David and Lu, Shuai and Jin, Zhi},
  booktitle = {Proceedings of the Twenty-seventh International Joint Conference on Artificial Intelligence (IJCAI 2018), Stockholm, Sweden, July},
  pages = {13--19},
  title = {Summarizing Source Code with Transferred API Knowledge},
  year = {2018}
}

@inproceedings{zhai2020cpc,
  author = {Zhai, Juan and Xu, Xiangzhe and Shi, Yu and Tao, Guanhong and Pan, Minxue and Ma, Shiqing and Xu, Lei and Zhang, Weifeng and Tan, Lin and Zhang, Xiangyu},
  booktitle = {Proceedings of the ACM/IEEE 42nd International Conference on Software Engineering},
  pages = {1359--1371},
  title = {CPC: Automatically Classifying and Propagating Natural Language Comments via Program Analysis},
  year = {2020}
}

@inproceedings{mu2023developer,
  author = {Mu, Fangwen and Chen, Xiao and Shi, Lin and Wang, Song and Wang, Qing},
  booktitle = {2023 IEEE/ACM 45th International Conference on Software Engineering (ICSE)},
  organization = {IEEE},
  pages = {768--780},
  title = {Developer-intent Driven Code Comment Generation},
  year = {2023}
}

@article{shor2003haven,
  author = {Shor, Peter W},
  journal = {Journal of the ACM (JACM)},
  number = {1},
  pages = {87--90},
  publisher = {ACM New York, NY, USA},
  title = {Why haven't more Quantum Algorithms been Found?},
  volume = {50},
  year = {2003}
}

@article{pearson1895vii,
  author = {Pearson, Karl},
  journal = {Proceedings of the Royal Society of London},
  number = {347-352},
  pages = {240--242},
  publisher = {The Royal Society London},
  title = {VII. Note on Regression and Inheritance in the Case of Two Parents},
  volume = {58},
  year = {1895}
}

@article{jaccard1912distribution,
  author = {Jaccard, Paul},
  journal = {New Phytologist},
  number = {2},
  pages = {37--50},
  publisher = {Wiley Online Library},
  title = {The Distribution of the Flora in the Alpine Zone. 1},
  volume = {11},
  year = {1912}
}

@article{zhao2020quantum,
  author = {Zhao, Jianjun},
  journal = {arXiv preprint arXiv:2007.07047},
  title = {Quantum Software Engineering: Landscapes and Horizons},
  year = {2020}
}

@article{reimers2019sentence,
  author = {Reimers, Nils and Gurevych, Iryna},
  journal = {arXiv preprint arXiv:1908.10084},
  title = {Sentence-BERT: Sentence Embeddings Using Siamese BERT-networks},
  year = {2019}
}

@book{nielsen2010quantum,
  author = {Nielsen, Michael A and Chuang, Isaac L},
  publisher = {Cambridge University Press},
  title = {Quantum Computation and Quantum Information},
  year = {2010}
}

@article{jayakumar2022quantum,
  author = {J., Abhijith and Adedoyin, Adetokunbo and Ambrosiano, John and Anisimov, Petr and Casper, William and Chennupati, Gopinath and Coffrin, Carleton and Djidjev, Hristo and Gunter, David and Karra, Satish and Lemons, Nathan and Lin, Shizeng and Malyzhenkov, Alexander and Mascarenas, David and Mniszewski, Susan and Nadiga, Balu and O’malley, Daniel and Oyen, Diane and Pakin, Scott and Prasad, Lakshman and Roberts, Randy and Romero, Phillip and Santhi, Nandakishore and Sinitsyn, Nikolai and Swart, Pieter J. and Wendelberger, James G. and Yoon, Boram and Zamora, Richard and Zhu, Wei and Eidenbenz, Stephan and B\"{a}rtschi, Andreas and Coles, Patrick J. and Vuffray, Marc and Lokhov, Andrey Y.},
  journal = {ACM Transactions on Quantum Computing},
  number = {4},
  pages = {18},
  title = {Quantum Algorithm Implementations for Beginners},
  volume = {3},
  year = {2022}
}

@article{serrano2022quantum,
  author = {Serrano, Manuel A and Cruz-Lemus, Jos{\'e} A and Perez-Castillo, Ricardo and Piattini, Mario},
  journal = {ACM Computing Surveys},
  number = {8},
  pages = {1--31},
  publisher = {ACM New York, NY},
  title = {Quantum Software Components and Platforms: Overview and Quality Assessment},
  volume = {55},
  year = {2022}
}

@inproceedings{wang2021qdiff,
  author = {Wang, Jiyuan and Zhang, Qian and Xu, Guoqing Harry and Kim, Miryung},
  booktitle = {2021 36th IEEE/ACM International Conference on Automated Software Engineering (ASE)},
  organization = {IEEE},
  pages = {692--704},
  title = {Qdiff: Differential Testing of Quantum Software Stacks},
  year = {2021}
}

@article{van2001pep,
  author = {Van Rossum, Guido and Warsaw, Barry and Coghlan, Nick},
  journal = {Python. Org},
  pages = {28},
  title = {PEP 8--Style Guide for Python Code},
  volume = {1565},
  year = {2001}
}

@article{goodger2001pep,
  author = {Goodger, David and van Rossum, Guido},
  journal = {Documentation, Python Software Foundation},
  title = {PEP 257--Docstring Conventions},
  year = {2001}
}

@inproceedings{svore2018q,
  author = {Svore, Krysta and Geller, Alan and Troyer, Matthias and Azariah, John and Granade, Christopher and Heim, Bettina and Kliuchnikov, Vadym and Mykhailova, Mariia and Paz, Andres and Roetteler, Martin},
  booktitle = {Proceedings of the Real World Domain Specific Languages Workshop 2018},
  pages = {1--10},
  title = {Q\# Enabling Scalable Quantum Computing and Development with a High-level dsl},
  year = {2018}
}

@misc{google2018cirq,
  author = {Google AI Quantum Team},
  howpublished = {\url{https://github.com/quantumlib/}},
  title = {Cirq},
  year = {2018}
}

@article{pezze20252030,
  author = {Pezz{\`e}, Mauro and Abrah{\~a}o, Silvia and Penzenstadler, Birgit and Poshyvanyk, Denys and Roychoudhury, Abhik and Yue, Tao},
  journal = {ACM Transactions on Software Engineering and Methodology},
  number = {5},
  pages = {1--55},
  publisher = {ACM New York, NY},
  title = {A 2030 Roadmap for Software Engineering},
  volume = {34},
  year = {2025}
}

@article{murillo2025quantum,
  author = {Murillo, Juan Manuel and Garcia-Alonso, Jose and Moguel, Enrique and Barzen, Johanna and Leymann, Frank and Ali, Shaukat and Yue, Tao and Arcaini, Paolo and P\'{e}rez-Castillo, Ricardo and Garc\'{\i}a-Rodr\'{\i}guez de Guzm\'{a}n, Ignacio and Piattini, Mario and Ruiz-Cort\'{e}s, Antonio and Brogi, Antonio and Zhao, Jianjun and Miranskyy, Andriy and Wimmer, Manuel},
  journal = {ACM Transactions on Software Engineering and Methodology},
  number = {5},
  pages = {1--48},
  publisher = {ACM New York, NY},
  title = {Quantum Software Engineering: Roadmap and Challenges Ahead},
  volume = {34},
  year = {2025}
}

@inproceedings{arafat2009comment,
  author = {Arafat, Oliver and Riehle, Dirk},
  booktitle = {2009 31st International Conference on Software Engineering-companion Volume},
  organization = {IEEE},
  pages = {195--198},
  title = {The Comment Density of Open Source Software Code},
  year = {2009}
}

@book{fenton2014software,
  author = {Fenton, Norman and Bieman, James},
  publisher = {CRC Press},
  title = {Software Metrics: A Rigorous and Practical Approach},
  year = {2014}
}

@book{schutze2008introduction,
  author = {Sch{\"u}tze, Hinrich and Manning, Christopher D and Raghavan, Prabhakar},
  publisher = {Cambridge University Press Cambridge},
  title = {Introduction to Information Retrieval},
  volume = {39},
  year = {2008}
}

@article{maaten2008visualizing,
  author = {Maaten, Laurens van der and Hinton, Geoffrey},
  journal = {Journal of Machine Learning Research},
  number = {Nov},
  pages = {2579--2605},
  title = {Visualizing Data Using T-SNE},
  volume = {9},
  year = {2008}
}

@article{huang2023comparative,
  author = {Huang, Yuan and Guo, Hanyang and Ding, Xi and Shu, Junhuai and Chen, Xiangping and Luo, Xiapu and Zheng, Zibin and Zhou, Xiaocong},
  journal = {ACM Transactions on Software Engineering and Methodology},
  number = {5},
  pages = {1--26},
  publisher = {ACM New York, NY},
  title = {A Comparative Study on Method Comment and Inline Comment},
  volume = {32},
  year = {2023}
}

@article{chen2021my,
  author = {Chen, Qiuyuan and Xia, Xin and Hu, Han and Lo, David and Li, Shanping},
  journal = {ACM Transactions on Software Engineering and Methodology (TOSEM)},
  number = {2},
  pages = {1--29},
  publisher = {ACM New York, NY, USA},
  title = {Why My Code Summarization Model does not Work: Code Comment Improvement with Category Prediction},
  volume = {30},
  year = {2021}
}

@article{leclair2019recommendations,
  author = {LeClair, Alexander and McMillan, Collin},
  journal = {arXiv preprint arXiv:1904.02660},
  title = {Recommendations for Datasets for Source Code Summarization},
  year = {2019}
}

@article{leite2025testing,
  author = {Leite Ramalho, Neilson Carlos and Amario de Souza, Higor and Lordello Chaim, Marcos},
  journal = {ACM Transactions on Software Engineering and Methodology},
  number = {5},
  pages = {1--46},
  publisher = {ACM New York, NY},
  title = {Testing and Debugging Quantum Programs: The Road to 2030},
  volume = {34},
  year = {2025}
}

@article{pontius2011death,
  author = {Pontius Jr, Robert Gilmore and Millones, Marco},
  journal = {International Journal of Remote Sensing},
  number = {15},
  pages = {4407--4429},
  publisher = {Taylor \& Francis},
  title = {Death to Kappa: Birth of Quantity Disagreement and Allocation Disagreement for Accuracy Assessment},
  volume = {32},
  year = {2011}
}

@inproceedings{devlin2019bert,
  author = {Devlin, Jacob and Chang, Ming-Wei and Lee, Kenton and Toutanova, Kristina},
  booktitle = {Proceedings of the 2019 Conference of the North American Chapter of the Association for Computational Linguistics: Human Language Technologies, Volume 1 (Long and Short Papers)},
  pages = {4171--4186},
  title = {BERT: Pre-training of Deep Bidirectional Transformers for Language Understanding},
  year = {2019}
}

@article{landis1977measurement,
  author = {Landis, J Richard and Koch, Gary G},
  journal = {Biometrics},
  pages = {159--174},
  publisher = {JSTOR},
  title = {The Measurement of Observer Agreement for Categorical Data},
  year = {1977}
}

@dataset{cc4q2025artifact,
  author       = {Zhou, Zenghui and Li, Yuechen and Cai, Yi and Wen, Jinlong and Yu, Xiaohan},
  title        = {Artifact of Code Comments for Quantum Software Development Kits: An Empirical Study on Qiskit},
  year         = {2025},
  month        = nov,
  publisher    = {Figshare},
  doi          = {10.6084/m9.figshare.30085657.v1},
  url          = {https://figshare.com/articles/dataset/Code_Comments_for_Quantum_Software_Development_Kits_An_Empirical_Study_on_Qiskit/30085657}
}

@article{SuMcMillan2024,
  title = {Distilled GPT for source code summarization},
  author = {Su, Chia-Yi and McMillan, Collin},
  year = 2024,
  journal = {Automated Software Engineering},
  volume = {31},
  number = {1},
  pages = {22},
  doi = {10.1007/s10515-024-00421-4},
}

@article{EvtikhievBogomolov2023,
  title = {Out of the BLEU: How should we assess quality of the Code Generation models?},
  author = {Evtikhiev, Mikhail and Bogomolov, Egor and Sokolov, Yaroslav and Bryksin, Timofey},
  year = 2023,
  journal = {Journal of Systems and Software},
  volume = {203},
  pages = {111741},
  doi = {10.1016/j.jss.2023.111741},
}

@inproceedings{RoyFakhoury2021,
  title = {Reassessing automatic evaluation metrics for code summarization tasks},
  booktitle = {Proceedings of the 29th ACM Joint Meeting on European Software Engineering Conference and Symposium on the Foundations of Software Engineering},
  author = {Roy, Devjeet and Fakhoury, Sarah and Arnaoudova, Venera},
  year = 2021,
  pages = {1105--1116},
  doi = {10.1145/3468264.3468588},
  isbn = {978-1-4503-8562-6},
}

@inproceedings{YuLi2025,
  title = {QuAInth: A Code Comment Approach for Application-Oriented Quantum Programs via N-Version LLMs},
  booktitle = {2025 25th International Conference on Software Quality, Reliability and Security (QRS)},
  author = {Yu, Xiaohan and Li, Yuechen and Wen, Jinlong and Cai, Kai-Yuan and Yin, Beibei},
  year = 2025,
  pages = {13--24},
  doi = {10.1109/QRS65678.2025.00013}
}

@inproceedings{IshimotoNakamura2024,
  title = {An Empirical Study on Self-Admitted Technical Debt in Quantum Software},
  booktitle = {2024 31st Asia-Pacific Software Engineering Conference},
  author = {Ishimoto, Yuta and Nakamura, Yuto and Katsube, Ryota and Sato, Naoto and Ogawa, Hideto and Kondo, Masanari and Kamei, Yasutaka and Ubayashi, Naoyasu},
  year = 2024,
  pages = {41--50},
  doi = {10.1109/APSEC65559.2024.00015}
}

@article{zhang2025quantum,
  title={Quantum Optimization for Software Engineering: A Survey},
  author={Zhang, Man and Li, Yuechen and Yue, Tao and Cai, Kai-Yuan},
  journal={arXiv preprint arXiv:2506.16878},
  year={2025}
}
